\documentclass[aps, prl, reprint,showpacs, showkeys,  amsmath, amsfonts, amssymb]{revtex4-1}
\usepackage{hyperref}
\usepackage{graphicx}
\usepackage{mathtools}
\usepackage[caption=false]{subfig}
\usepackage{tikz}
\usepackage{braket}

\def\beq{\begin{equation}}
\def\eeq{\end{equation}}
\def\beqa{\begin{eqnarray}}
\def\eeqa{\end{eqnarray}}
\def\bsub{\begin{subequations}}
	\def\esub{\end{subequations}}
\def\bal{\begin{align}}
\def\eal{\end{align}}

\newcommand\scale{0.4}

\newcommand\bs{0.15cm}

\def\lb{\rotatebox[origin=c]{-45}{\textemdash}}
\def\rb{\rotatebox[origin=c]{45}{\textemdash}}

\newcommand{\rstatevar}[2]
{
	\begin{tikzpicture}[scale=0.4,baseline=0.15cm]
	\node at (1,1) {#1};
	\node at (4,0) {#2};
	\node at (2.5,0.5) {\ldots};
	\draw[black, thin] (1,0) -- (2,1);
	\draw[black, thin] (3,0) -- (4,1);
	\end{tikzpicture}
}

\newcommand{\lstatevar}[2]
{
	\begin{tikzpicture}[scale=0.4,baseline=0.15cm]
	\node at (1,0) {#1};
	\node at (4,1) {#2};
	\node at (2.5,0.5) {\ldots};
	\draw[black, thin] (1,1) -- (2,0);
	\draw[black, thin] (3,1) -- (4,0);
	\end{tikzpicture}
}

\begin{document}
	
\title{Quantum many-body scars in transverse field Ising ladders and beyond}
\date{\today}
\author{Bart \surname{van Voorden}}
\email[Email: ]{B.A.vanVoorden@uva.nl}
\author{Ji\v{r}\'{i} \surname{Min\'{a}\v{r}}}
\author{Kareljan \surname{Schoutens}}
\affiliation{Institute for Theoretical Physics, University of Amsterdam,
	Science Park 904, 1098 XH Amsterdam, the Netherlands}

\begin{abstract}

We identify quantum many-body scars in the transverse field quantum Ising model on a ladder. We make explicit how the corresponding (mid spectrum, low entanglement) many-body eigenstates can be approximated by injecting quasi-particle excitations into an exact, zero-energy eigenstate, which is of valence bond solid type. Next, we present a systematic construction of product states characterized, in the limit of a weak transverse field, by a sharply peaked local density of states. We describe how the construction of these `peak states’ generalizes to arbitrary dimension and show that on the ladder their number scales with system size as the square of the golden ratio.

\end{abstract}

\maketitle

\textit{Introduction.}  Understanding of non-equilibrium dynamics and thermalization is at the forefront of research on quantum many-body systems. These developments led to the formulation of eigenstate thermalization hypothesis (ETH)~\cite{Deutsch1991, Srednicki1994,Rigol2008}, predicting fast thermalization following a quench from a generic many-body state. A number of exceptions to this behaviour have been identified, namely integrable~\cite{Sutherland2004,Takahashi_2005_Book} and many-body localized systems~\cite{Gornyi2005, Basko2006,Serbyn_2013_PRL,Huse_2014_PRB, Nandkishore2014,Imbrie_2016_JStatPhys,Imbrie_2017_AnnPhys}, which both preclude thermalization due to a number of conserved charges. A recent observation of non-thermalizing behaviour in a chain of Rydberg atoms \cite{Bernien2017} described by a so-called PXP Hamiltonian~\cite{Lesanovsky_2012_PRA}, has been interpreted in terms of quantum many-body scars (QMBS)~\cite{Turner2018, Turner2018a}, which have been named in analogy to early works on single particle quantum scars~\cite{Heller_1984_PRL,Kaplan_1998_AnnPhys,Kaplan_1999}. QMBS are an example of weak ETH-breaking, i.e. lack of thermalization for a limited set of typically weakly entangled initial states. This observation led to a number of works including further studies on the PXP-model~\cite{Choi,Lin2018,Iadecola2019,Surace_2019,Lin_2019,Bull_2020,Mark_2019,Lin_2020,Yang_2020_PRL},  constrained ~\cite{Ho2018a, Bull2019,Pancotti_2019,Roy_2019} and topological ~\cite{Ok2019} Hamiltonians featuring non-thermal states, studies of the role of integrability~\cite{Khemani2018}, quantum chaos~\cite{Hallam_2019_NatComm, Jansen_2019_PRB, Moudgalya_2019_PRB, Wilming_2019_PRL, Michailidis_2020_PRX,Andreev_2019,Werman_2020} or fragmentation of Hilbert space~\cite{DeTomasi_2019_PRB, Khemani_2019, Sala_2020_PRX,Karpov_2020}. Simultaneously, QMBS have been described in a range of models, including the AKLT model~\cite{Moudgalya_2018_PRB,Moudgalya2018,Shiraishi_2019_JStatMech,Mark_2020,Moudgalya_2020}, spin chains~\cite{Chattopadhyay2019, Schecter2019, Iadecola_2019_PRL,Shibata_2019} and arrays~\cite{Lee_2020}, boson~\cite{Chen_2020_PRA} and spin-boson~\cite{Sinha_2019,Villasenor_2020}, and driven systems~\cite{Pai2019, Pizzi_2019,Sugiura_2019,Zhao2020,Mukherjee_2020,Jen_2020_PRR,Fan_2020_PRA}.

Interestingly and to the best of our knowledge, the paradigmatic model of quantum magnetism, the quantum Ising model, has not been analysed from the perspective of QMBS beyond a chain~\cite{Yang_2020_PRL}. Here we note that Refs.~\cite{James_2019_PRL, Robinson_2019_PRB} investigated non-thermal behaviour in the longitudinal field Ising model (cf. also~\cite{Bukva_2019}) following a quench, where the dynamics have been interpreted in terms of meson quasiparticles~\cite{Kormos_2017_NatPhys} while Ref.~\cite{Iadecola_2020_PRB} studied a deformed Ising Hamiltonian in one spatial dimension.
 
In this work we analyse the transverse field Ising model on a ladder. We find a number of initial product states which feature quasiperiodic revivals in the autocorrelation function, signalling non-thermal behaviour. We describe a systematic construction of these states, which in some cases extend from the ladder to higher dimensions. We further provide a number of analytical results, namely an expression for a zero energy transverse field independent eigenstate which we identify as a valence bond solid~\cite{Affleck1987} and which serves as a starting point for a systematic construction of scarred eigenstates
\footnote{
After the completion of this work, we became aware of Ref.~\cite{Lin_2020} which have identified QMBS in two-dimensional PXP-model linking them also to the the VBS.}.
Additional analytical results are obtained for the energy separation between the scarred eigenstates, the degeneracies of the zero transverse field manifolds, and the number of ETH-breaking `peak' states, which scales as the golden ratio. These results constitute a direct experimental recipe for QMBS in Ising models which have been already realized with Rydberg quantum simulators~\cite{Schauss_2015_Science, Labuhn_2016_Nature,Browaeys_2020_NatPhys} including the ladder geometry~\cite{deLeseleuc_2019_Science}. 

\textit{Model.}---We consider the transverse field Ising model on a $L\times2$ ladder, with $L$ even, and Hamiltonian 
\begin{equation}\label{Hamiltonian}
	H = H_z + H_x = \sum_{\langle i,j \rangle} \sigma^z_{i} \sigma^z_{j} + h_x \sum_i \sigma^x_{i},
\end{equation}
where $\langle i,j \rangle$ denotes nearest neighbours, $\sigma^z_{i} = c^\dagger_i c^{\phantom{\dagger}}_i - c^{\phantom{\dagger}}_i c_i^\dagger$, $\sigma^x_{i} = c^\dagger_i + c^{\phantom{\dagger}}_i$ are the Pauli matrices expressed in terms of the hard-core bosonic operators with the usual commutation relations $[c_i,c_j^\dag]=\delta_{ij}( 1- 2 c_i^\dag c_i)$. We further assume periodic boundary conditions $c_{x+L,y}^\dagger=c_{x,y}^\dagger$, $c_{x,y+2}^\dagger=c_{x,y}^\dagger$ where we have introduced real-space coordinates of each site $i = (x,y)$. The many-body basis states of the Hamiltonian are  
\begin{equation}\label{basis_definition}
\small \left|b \right\rangle = \left|   \begin{array}{cccc}  n_{0,1} & n_{1,1} & \ldots & n_{L-1,1}\\ n_{0,0} &n_{1,0} &\ldots & n_{L-1,0} \end{array} \right\rangle, 
\end{equation}
where $n_{x,y} \in \{0,1\}$ is the occupation number.
 
For $h_x=0$ the eigenvectors of $H$ are the basis states~(\ref{basis_definition}). Their energy is determined by the difference in the number of equal and unequal neighbouring pairs. Each pair $\left\langle i,j\right\rangle$ with $n_i=n_j$ ($n_i\neq n_j$) adds $E=1$ ($E=-1$) to the energy. 
Therefore, the total energy is given by the difference in the number of equal and unequal neighbour pairs. This leads to all energy levels being degenerate~\cite{Suppl}. Flipping a single spin in a given configuration changes the sign of the energy contribution of all neighbouring sites, so that the energy difference between the degenerate manifolds is $|\Delta E|=4$ with the exception of the lowest (highest) and first (de-)excited ones for which $|\Delta E|=8$. The highest (lowest) energy state of $E=4L$ ($E=-4L$) is reached when all sites have the same $n_{x,y}$ (all neighbours of each site have opposite $n_{x,y}$). Defining the potential $V_i \equiv \Braket{b_i | H_z |b_i}$, the above implies $V_{i}=4m_i$ for integer $|m_i| \leq L$~\cite{Suppl}.


We now consider $h_x \neq 0$. We note that $\left\langle b_i|H_x |b_j \right\rangle \neq 0$ only if $\left|b_i\right\rangle$ and $\left| b_j\right\rangle$ differ by a single spin flip. When $V_i \neq V_j$, the perturbative correction to the energies $E_i$, $E_j$ due to the matrix element $\langle b_i | H_x | b_j \rangle$ is of order $h_x^2$ so that it is strongly suppressed when $h_x\ll 1$. In contrast, basis states with $V_i=V_j$ will hybridize under the perturbation. Consequently, considering a perturbation up to first order in $h_x$ is equivalent to using the projected Hamiltonian
\begin{equation}\label{Hamiltonian_PXP}
H' = H_z + H'_x = \sum_{\langle i,j \rangle} \sigma^z_{i} \sigma^z_{j} + h_x \sum_i P\sigma^x_{i}P,
\end{equation}
where $P$ acting on $\left| b_i\right\rangle$ projects it on all $\left| b_j\right\rangle$ with $V_j=V_i$~\footnote{Despite the similarity, the Hamiltonian $H'$ is not equal to the PXP-model due to the difference in the projection operators $P$.}. 

\begin{figure}
	\centering
	\includegraphics[width=\linewidth]{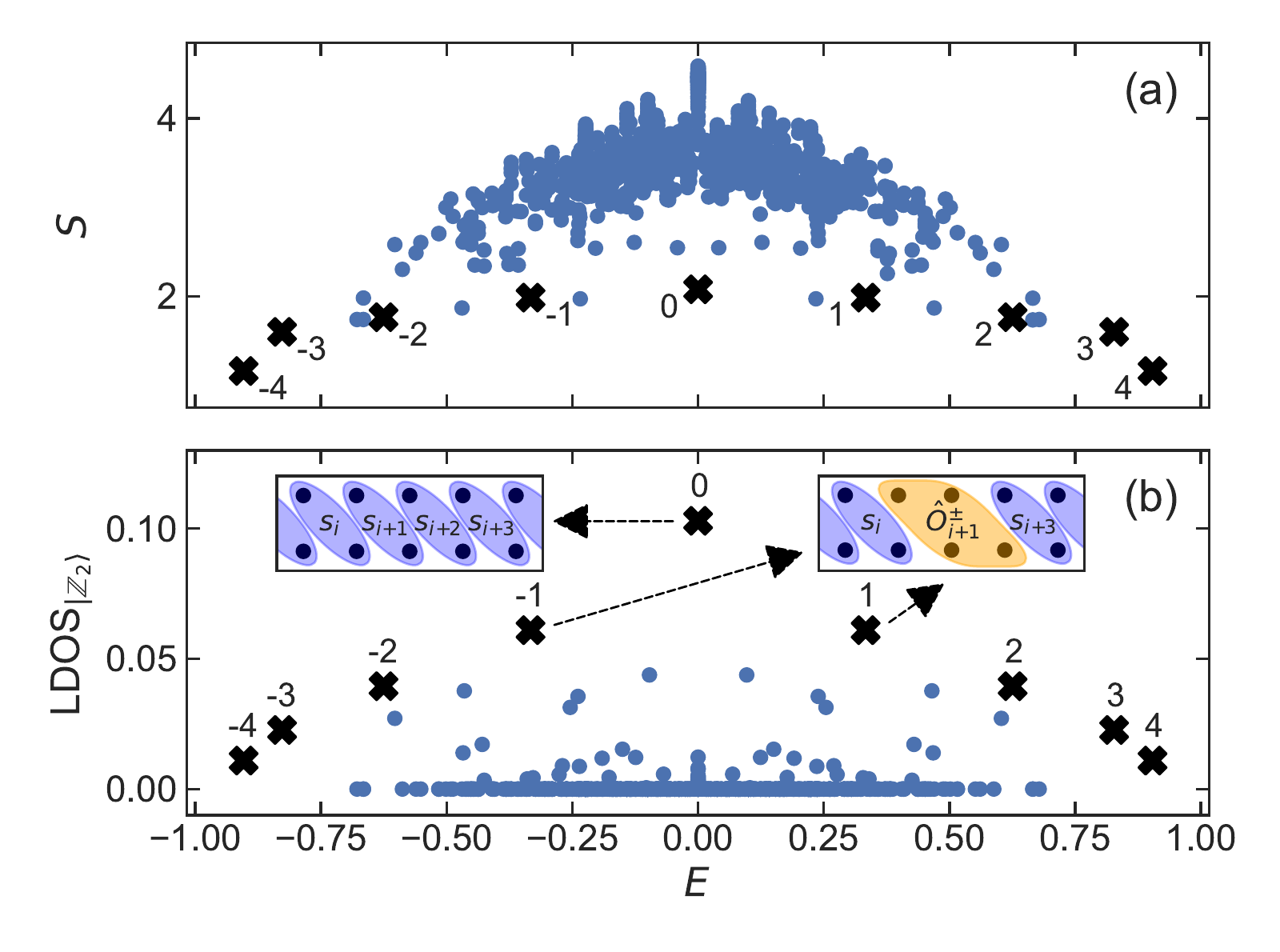}
	\caption{(a) The entanglement entropy and (b) the LDOS$_{\left|\mathbb{Z}_2\right\rangle}$ of the eigenvectors of $H'$, Eq.~(\ref{Hamiltonian_PXP}), in the vicinity of $E=0$ for $L=8$, $h_x=0.1$ and $k_x=k_y=0$.	
	The numbered crosses indicate the $\left| v_{\text{\scriptsize{QMBS}},n}\right\rangle$. The left (right) inset in (b) shows the approximation  $\left| \psi_{E=0}'\right\rangle$ ($\left|\psi_{\text{\scriptsize{SMA}},\pm 1}' \right\rangle$), Eq.~(\ref{exactzero}) [(\ref{SMA})].}
		\label{fig:overlapz2_entropy_L8_v2}
\end{figure}

\textit{Scars.}---It has been argued that QMBS typically correspond to low entanglement entropy~(EE) states~\cite{Turner2018a,Ho2018a,Moudgalya2018}. To this end we consider the (second R\'{e}nyi) EE $S = - \ln\left(\text{Tr}(\rho_A^2)\right)$ with $\rho_A$ the density matrix of a subsystem $A$, which we take to be the half-ladder $L_x=L/2$ unless stated otherwise. The EE spectrum for the $E=0$ manifold perturbed by the transverse field shows a band of low EE states $\left| v_{\text{\scriptsize{QMBS}},n}\right\rangle$ with $n=-L/2,\ldots,L/2$, characteristic of a QMBS~\cite{Turner2018}, see the black crosses in Fig.~\ref{fig:overlapz2_entropy_L8_v2}a.

Additionally, we consider the local density of states of a state $\left| \psi_0\right\rangle$,
\begin{equation}\label{LDOS}
\text{LDOS}_{\left| \psi_0\right\rangle}(E) = \sum_j \left|\left\langle v_j| \psi_0 \right\rangle\right|^2 \delta(E-E_j),
\end{equation}
where $\left| v_i\right\rangle$ are the eigenvectors of the Hamiltonian. Using $H'$, it can be seen from Fig.~\ref{fig:overlapz2_entropy_L8_v2}b, that the low EE eigenvectors feature high LDOS for specific product states, namely the $\ket{\mathbb{Z}_2}$ states, a situation analogous to the PXP model. Here ${\rm LDOS}_{\ket{\mathbb{Z}_2}}$ is identical for $\ket{\mathbb{Z}_2}$ being
\begin{equation}\label{Z_2_1}
\small \left|\mathbb{Z}_{2}^{\text{rung}} \right\rangle = \left|  \scriptsize{ \begin{array}{ccccc}  1 & 0 & 1 & 0 & \cdots \\ 1 & 0 & 1 & 0  &\cdots \end{array}} \right\rangle, 
\left|\mathbb{Z}_{2}^{\text{leg}} \right\rangle = \left|  \scriptsize{ \begin{array}{ccccc}  1 & 1 & 1 & 1 & \cdots \\ 0 & 0 & 0 & 0  &\cdots \end{array} }\right\rangle
\end{equation}
and their translations $T_x \left|\mathbb{Z}_2^{\text{rung}} \right\rangle$ and $T_y \left|\mathbb{Z}_2^{\text{leg}} \right\rangle$, where $T_{x,y}$ translates the state by one site in the $x,y$ direction. Denoting by $k_{x,y}$ the eigenvalues of $T_{x,y}$, it is convenient to work in the $k_x=k_y=0$ momentum sector of the Hamiltonian, and therefore we consider either $1/\sqrt{2}(1+ T_x)\left|\mathbb{Z}_2^{\text{rung}} \right\rangle$ or $1/\sqrt{2}(1+ T_y)\left|\mathbb{Z}_2^{\text{leg}} \right\rangle$. 
 
 \begin{figure}
 	\centering
 	\includegraphics[width=\linewidth]{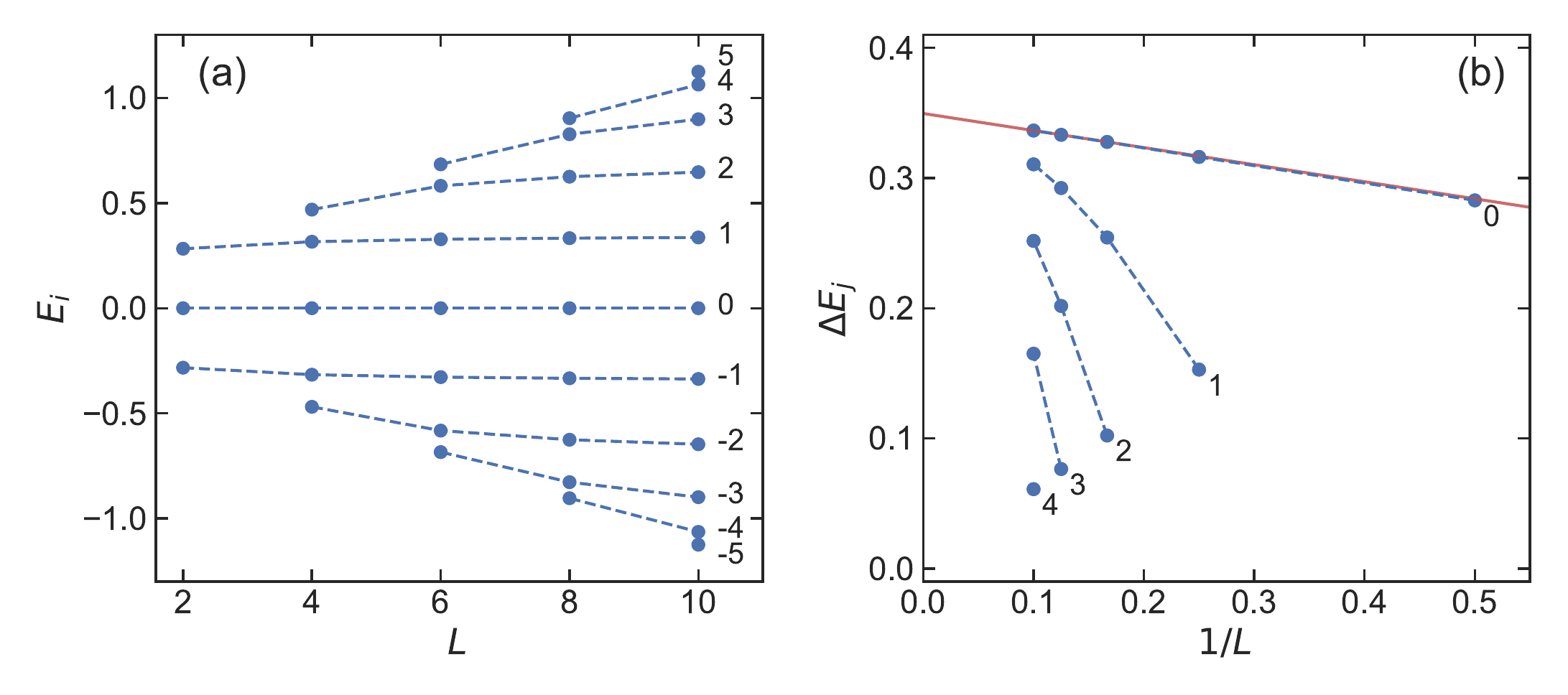}
 	\caption{(a) The energies $E_i$ and (b) energy differences $\Delta E_j$ of $\left| v_{\text{\scriptsize{QMBS}},n}\right\rangle$ as function of system size for $h_x=0.1$. The numbering of the states corresponds to Fig.~\ref{fig:overlapz2_entropy_L8_v2}. The red line is a linear fit for $\Delta E_0$. }
 	\label{fig:finite_size_scaling_energies_combined}
 \end{figure}
 
We note from Fig.~\ref{fig:overlapz2_entropy_L8_v2}b that the $\left| v_{\text{\scriptsize{QMBS}},n}\right\rangle$ are non-degenerate, except for the one at $E=0$, and have momentum $k_x=k_y=0$. Additionally, the energy separations between the special eigenstates are approximately equal. In  Fig.~\ref{fig:finite_size_scaling_energies_combined}a we show the respective energies, which we enumerate as $E_{-L/2},\ldots, E_{L/2}$ and the energy differences $\Delta E_j = E_{j+1}-E_j$ between two consecutive energies are shown in Fig.~\ref{fig:finite_size_scaling_energies_combined}b (we note that $\Delta E_{-j} = \Delta E_{j-1}$ for $j>0$ due to the spectral inversion symmetry).  As can be seen, the energy differences tend to become more equal for increasing $L$. A linear fit to $\Delta E_0$ shows that in the limit $L\rightarrow\infty$ the energy difference becomes $\Delta E_0 \approx 3.49h_x$. 

\begin{figure}
  	\centering
  	\includegraphics[width=\linewidth]{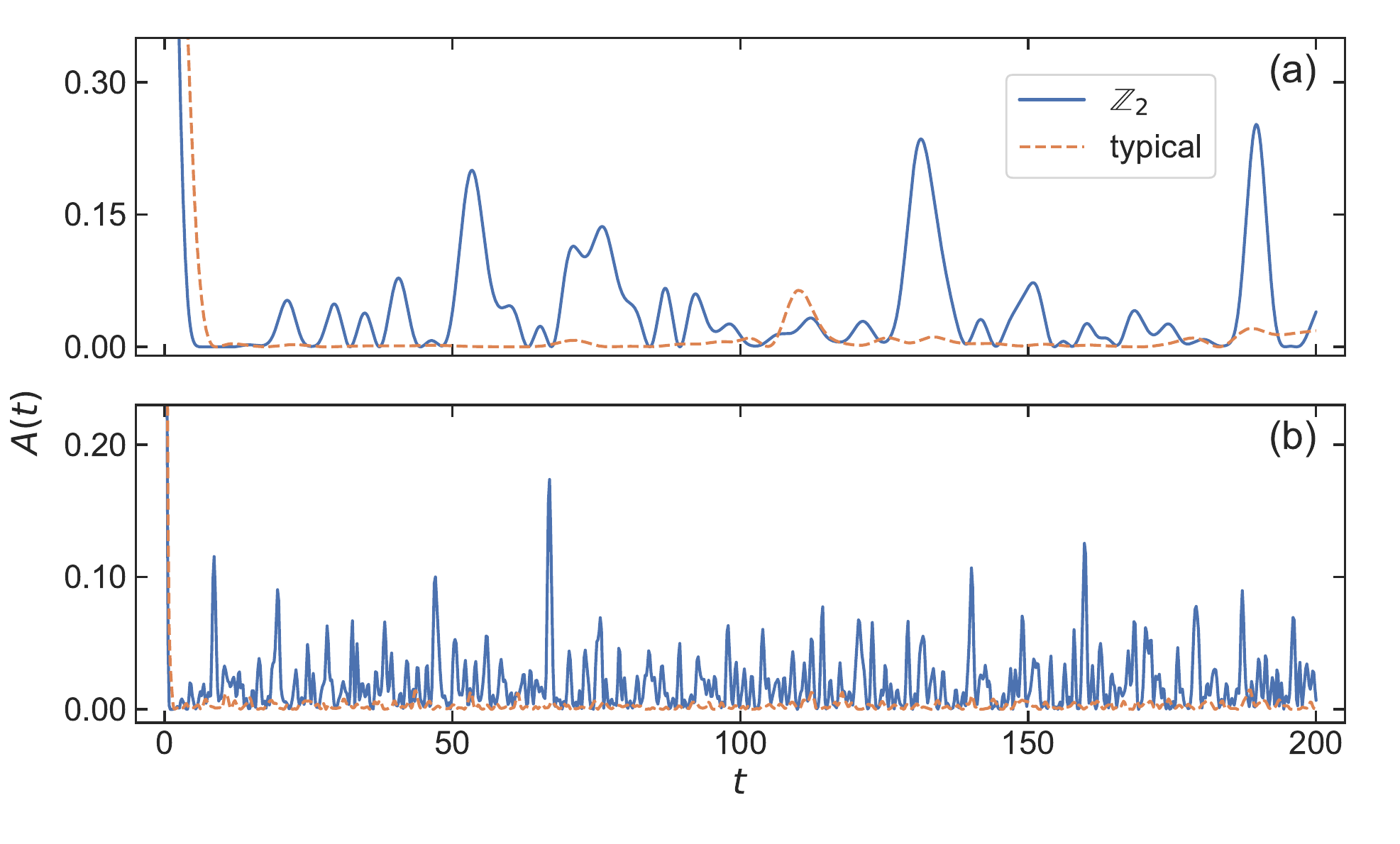}
  	\caption{
  	 The autocorrelation Eq.~(\ref{autocor}) for a $\left|\mathbb{Z}_2\right\rangle$ (solid, blue) and a typical (dashed, orange) states for $L=8$ with (a) $h_x=0.1$ and (b) $h_x=0.7$.
  	}
  	\label{fig:time_evolution_Z2_other_L8_combinedhx}
  \end{figure}

As a direct consequence of the approximately equal energy difference between the peaks in $\text{LDOS}_{\left|\mathbb{Z}_2 \right\rangle}$, initializing the system in $\left| \psi_0\right\rangle = \left|\mathbb{Z}_2\right\rangle$ leads to ETH-breaking behaviour. To characterize the ensuing quench dynamics, we consider the Loschmidt echo or autocorrelation~\cite{Alhambra_2019},
  \begin{equation}\label{autocor}
  A_{\left| \psi_0\right\rangle}(t) = \left|\left\langle \psi(t)| \psi_0 \right\rangle\right|^2  = \left|\left\langle \psi_0| e^{-i H t}|\psi_0\right\rangle\right|^2,
  \end{equation}
  which is related to $\text{LDOS}_{\left| \psi_0\right\rangle}$ by a Fourier transform~\cite{Heller_1984_PRL}. Because of the special properties of $\text{LDOS}_{\left|\mathbb{Z}_2 \right\rangle}$, the autocorrelation shows revivals of the initial wave function, see Fig.~\ref{fig:time_evolution_Z2_other_L8_combinedhx}a,b, where the full Hamiltonian~(\ref{Hamiltonian}) was used with $h_x=0.1$ and $h_x=0.7$. Here $A_{\left|\mathbb{Z}_2 \right\rangle}(t)$ shows revivals, in contrast to the autocorrelation of a typical basis state~\cite{Suppl} (for simplicity, here and in the following we omit the explicit normalization unless stated otherwise)
  \begin{equation}\label{psityp}
  \left|\psi_{\text{typ}}\right\rangle = \sum_{i=0}^{L-1}\sum_{j=0}^1 T_x^{i}\, T_y^{j} \small{ \left| \begin{array}{cccccccc} 1 & 1 & 1 & 0 & 1 & 1 & 0 & 0 \\  1 & 0 & 1 & 0 & 1 & 0 & 0 & 0\end{array} \right\rangle}.
  \end{equation}
    To compare the response from different initial states, we calculate the average autocorrelation $\langle A \rangle=\lim_{\tau\rightarrow \infty} \tau^{-1}\int_{t_0}^\tau A(t)\,dt$~\footnote{We take $t_0>0$ such that the initial decay of $A(t)$ is not included in the average and $\tau$ big enough such that $\left\langle A\right\rangle$ is $\tau$-independent~\cite{Suppl}.}. We observe that $\braket{A_{\ket{\mathbb{Z}_2}}}$ also remains significantly higher than that of a typical state for $h_x$ beyond the perturbative regime and far from the integrable point at $h_x=0$, hinting towards the robustness of the ETH-breaking in the Ising ladder. The numerical results up to $L=8$ suggest that $\lim_{L \to \infty}\braket{A_{\ket{{\mathbb Z}_2}}}/\braket{A_{\ket{\psi_{\rm typ}}}} \gg 1$, while $\lim_{L \to \infty}\braket{A_{\ket{{\mathbb Z}_2}}} \approx 0$~\cite{Suppl}.
  
\textit{Analytical construction of QMBS.}---
We identified an $h_x$-independent $E=0$ eigenstate 
\begin{equation} \label{exactzero}
\left| \psi_{E=0}'\right\rangle = \prod_{i}^{L} s_i\left|\emptyset\right\rangle \equiv \prod_{i}^{L}\left(c^\dagger_{i,1} - c^\dagger_{i+1,0}\right)\left|\emptyset\right\rangle 
\end{equation}
where $\left|\emptyset\right\rangle = \scriptsize \left| \begin{array}{cc} 0 & \cdots \\ 0 & \cdots \end{array} \right\rangle$ is the vacuum state~\cite{Suppl}. $\ket{\psi'_{E=0}}$ is a product state of singlets aligned diagonally in the ladder, see the inset in Fig.~\ref{fig:overlapz2_entropy_L8_v2}b, and thus corresponds to a valence bond solid~(VBS) crystalline order. The knowledge of $|\psi'_{E=0}\rangle$ allows for a systematic construction of the descendant scarred eigenstates $\left| v_{\text{\scriptsize{QMBS}},n}\right\rangle$, indicated by crosses in Fig.~\ref{fig:overlapz2_entropy_L8_v2}. This is achieved by repeated action of a local operator creating quasiparticle excitations on top of $|\psi'_{E=0}\rangle$, a method termed the single mode approximation~(SMA) and applied to the PXP model using matrix product states~\cite{Lin2018}. Motivated by the form of $|\psi'_{E=0}\rangle$, we consider excitations by locally replacing a size $2\times2$ plaquette with an exact $E_\pm= \pm2\sqrt{2}h_x$ eigenstate of $H'$~(\ref{Hamiltonian_PXP}) with $L=2$. This eigenstate is $\hat{O}^{\pm}\left| \emptyset\right\rangle$ with
\begin{align}
\hat{O}^{\pm} = &\frac{1}{4} (1+Q) \left( c^\dagger_{0,0} + c^\dagger_{0,1} + c^\dagger_{1,0} + c^\dagger_{1,1}\right) \label{SMA} \\
\nonumber & \pm  \frac{1}{2\sqrt{2}} \left(c^\dagger_{0,0}c^\dagger_{0,1} + c^\dagger_{1,0}c^\dagger_{1,1} + c^\dagger_{0,0}c^\dagger_{1,0} + c^\dagger_{0,1}c^\dagger_{1,1} \right),
\end{align}
where $Q$ is the particle-hole inversion operator ($n_{x,y} \rightarrow 1-n_{x,y}$). It has an overlap of $1- h_x^2/32 +O(h_x^4)$ with the eigenvector of $H$ with $E_\pm= \pm2\sqrt{2}h_x \mp h_x^3/\sqrt{32} + O(h_x^5)$. However, in order to keep the correct energy, no singlet is allowed to be broken and therefore the excitation operator~(\ref{SMA}) has to be placed diagonally on the ladder. Consequently, we define the operator $\hat{O}^{\pm}_{j}$ as placing an excitation according to Eq.~(\ref{SMA}) on the sites at $(x,y) = (j,1), (j+1,1), (j+1,0)$ and $(j+2,0)$, see the inset in Fig.~\ref{fig:overlapz2_entropy_L8_v2}b, such that
\begin{equation}\label{psiSMA}
	\left|\psi_{\text{\scriptsize{SMA}},\pm 1}' \right\rangle = \sum_j \hat{O}^{\pm}_{j}\, \prod_{\mathclap{\scriptsize{ i\neq j,j+1}}} s_i \left|\emptyset\right\rangle\,.
\end{equation}

This method can be continued to create approximations to all $\left| v_{\text{\scriptsize{QMBS}},n}\right\rangle$ by adding more local excitations as
\begin{equation}\label{SMAn}
\left|\psi_{\text{\scriptsize{SMA,n}}}' \right\rangle = \sum_{k_1 \ldots k_n} \hat{O}^{\pm}_{k_1}\ldots \hat{O}^{\pm}_{k_n}\, \prod_{i}\, s_i \left|\emptyset\right\rangle,
\end{equation}
 where the sign in $\hat{O}^{\pm}$ corresponds to the sign of $n$, cf. Fig.~\ref{fig:overlapz2_entropy_L8_v2}b. The excitations are not allowed to overlap, so the indices have to obey $i,k_j \neq k_m,k_{m+1} \,\forall \, i,j,m$.
 
\begin{figure}
	\centering
	\includegraphics[width=.9\linewidth]{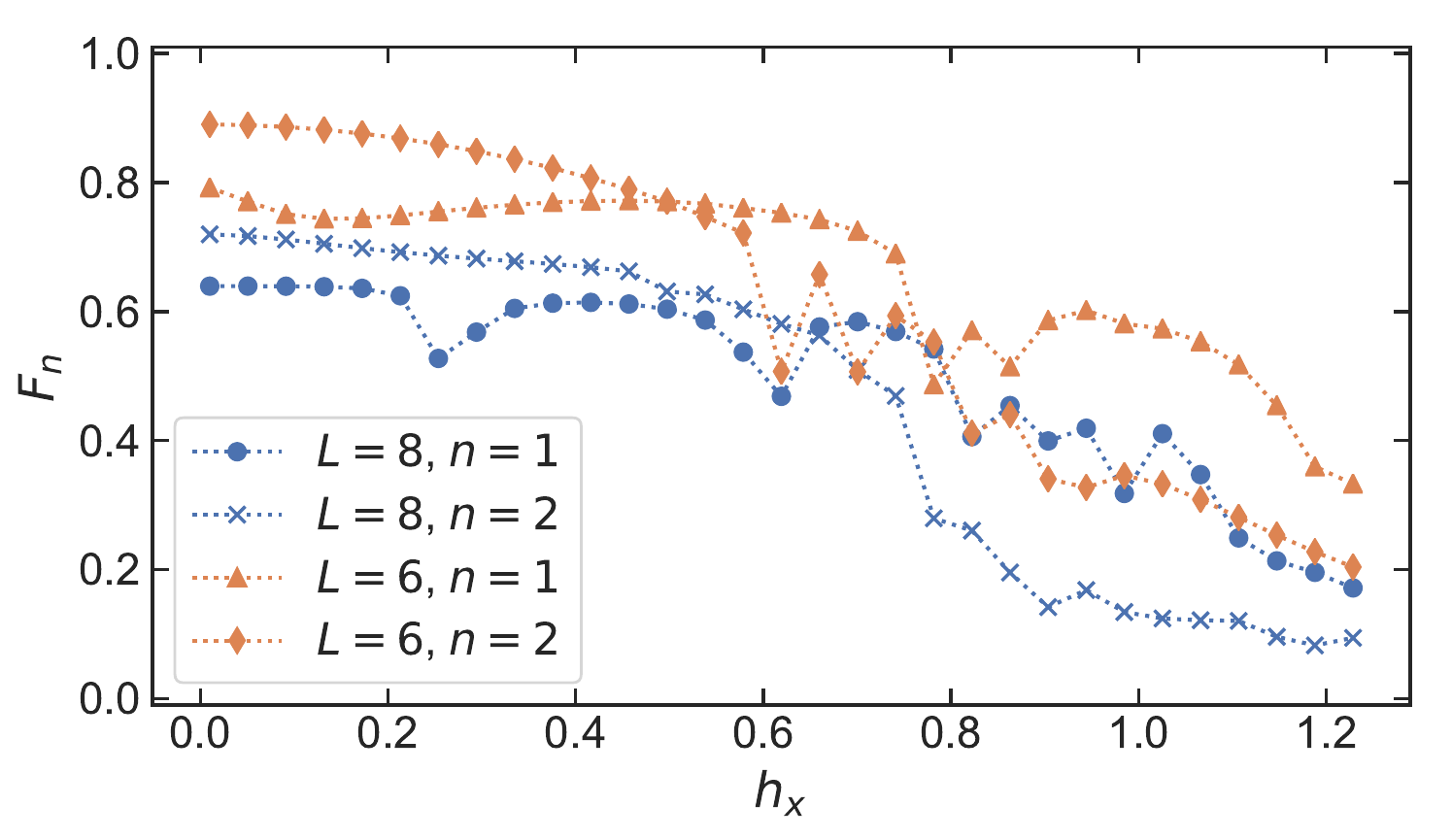}
	\caption{Fidelity $F_n$ of the first two SMA states vs. $h_x$ for $L=8$ (blue) and $L=6$ (orange).}
	\label{fig:overlapSMA_all}
\end{figure}

 The $\left| v_{\text{\scriptsize{QMBS}},n}\right\rangle$ states are symmetric under the reflection of the $x$-coordinates $(c^\dagger_{x,y}\rightarrow c^\dagger_{L-x,y})$, but the states~(\ref{exactzero}) and~(\ref{SMAn}) are not, because both the singlets and the excitations are placed diagonally on the ladder. In order to symmetrize these states we take the normalized superpositions 
\begin{subequations}
\label{eq:SMA}
 \begin{align}
 \label{exactzero_sym}
 \left| \psi_{E=0}\right\rangle &=  \sqrt{\frac{1}{2+2^{3-L}}}\,(1+R) \left| \psi_{E=0}'\right\rangle\\
 \left| \psi_{\text{\scriptsize{SMA, n}}}\right\rangle &= \mathcal{N}_n\,(1+R) \left| \psi_{\text{\scriptsize{SMA, n}}}'\right\rangle
 \end{align}
\end{subequations}
 where $R$ is reflection operator, $R:c^\dagger_{x,y} \rightarrow c^\dagger_{L-x,y}$, and $\mathcal{N}_n$ denotes the $n$-dependent normalization, e.g. $\mathcal{N}_{\pm1}= 1/\sqrt{2+2^{3-L}L}$~\cite{Suppl}.
 The state $\left| \psi_{E=0}\right\rangle$ has an overlap of $\left|\left\langle \psi_{E=0} | \mathbb{Z}_2\right\rangle\right|^2 = 1/(1+2^{L-2})$, which is high compared to a typical state, but low compared to the special states shown in Fig.~\ref{fig:overlapz2_entropy_L8_v2}b, making $\left| \psi_{E=0}\right\rangle$ a rather poor approximation to the exact eigenvector. In contrast, the overlap $\left|\left\langle \psi_{\text{SMA},1} | \mathbb{Z}_2\right\rangle\right|^2 = L/(2L+2^{L-1})$ corresponds well to that in Fig.~\ref{fig:overlapz2_entropy_L8_v2}b and the fidelity $F_{n}=\left|\left\langle v_{\text{\scriptsize{QMBS}},n} | \psi_{\text{\scriptsize{SMA}},n} \right\rangle\right|^2$ with $n=1$ and $n=2$, plotted in Fig.~\ref{fig:overlapSMA_all} for a range of $h_x$-values and for two different lengths $L$, shows that they are a reasonable approximation to the real eigenstates. The quality of the approximation decreases with increasing $h_x$ and $L$, which is expected due to the $h_x$-dependent reduced fidelity of the $\hat{O}$-operator to the exact plaquette eigenstates. We note a similar decrease of fidelity also appears in the SMA applied to the PXP model~\cite{Lin2018}. In~\cite{Suppl} we present a systematic construction, using the forward scattering approximation (FSA), of  a series of states $\left| w_{(j)}\right\rangle$, with $\left| w_{(1)}\right\rangle=\left| \psi'_{\text{\scriptsize{SMA}},1}\right\rangle$, converging on an eigenstate $\left| v'_{\text{\scriptsize{QMBS}},n}\right\rangle$ of $H'$, where up to normalization, $\left| v_{\text{\scriptsize{QMBS}},n}\right\rangle = (1+R)\left| v'_{\text{\scriptsize{QMBS}},n}\right\rangle$.

It follows directly from the structure of $\left| \psi_{E=0}'\right\rangle$, Eq. (\ref{exactzero}), that it has a constant half-ladder EE of $S=2\ln(2)$. It is independent of $L$ and of the size of the subsystem $A$, because the edges of the subsystem always have to cut through two singlet bonds, one at each edge (the leg entanglement entropy is $L\ln(2)$, because the subsystem cuts through all $L$ singlets). For $\left| \psi_{E=0}\right\rangle$ the half-chain entanglement entropy is
\begin{equation}\label{exactzero_entropy}
S_{\left| \psi_{E=0}\right\rangle} = -\ln\left(\frac{c+w(6+w)}{8(1+w)^2}\right),
\end{equation}
 with $w=2^{L-2}$, and $c=4$ ($c=1$) for $L/2$ even (odd)~\cite{Suppl}. It reaches its maximum in the thermodynamic limit, $\lim_{L \to \infty} S = 3 \ln(2)$ for both $L/2$ odd and even. We thus confirm the expected result, namely that, due to the product state nature of the underlying VBS states, both $\ket{\psi'_{E=0}}$ and $\ket{\psi_{E=0}}$ feature area-law EE in contrast to the typical eigenstates as reflected in Fig.~\ref{fig:overlapz2_entropy_L8_v2}a.

\textit{Other ETH-breaking states.}---In addition to the paradigmatic $\left| \mathbb{Z}_2\right\rangle$ states, we were able to identify a family of ETH-breaking product states in the $h_x\ll1$ limit, which offer a distinct experimental probe. They are characterized by a number of sharp peaks in the LDOS as we now describe. 

\begin{figure}
	\centering
	\includegraphics[width=\linewidth]{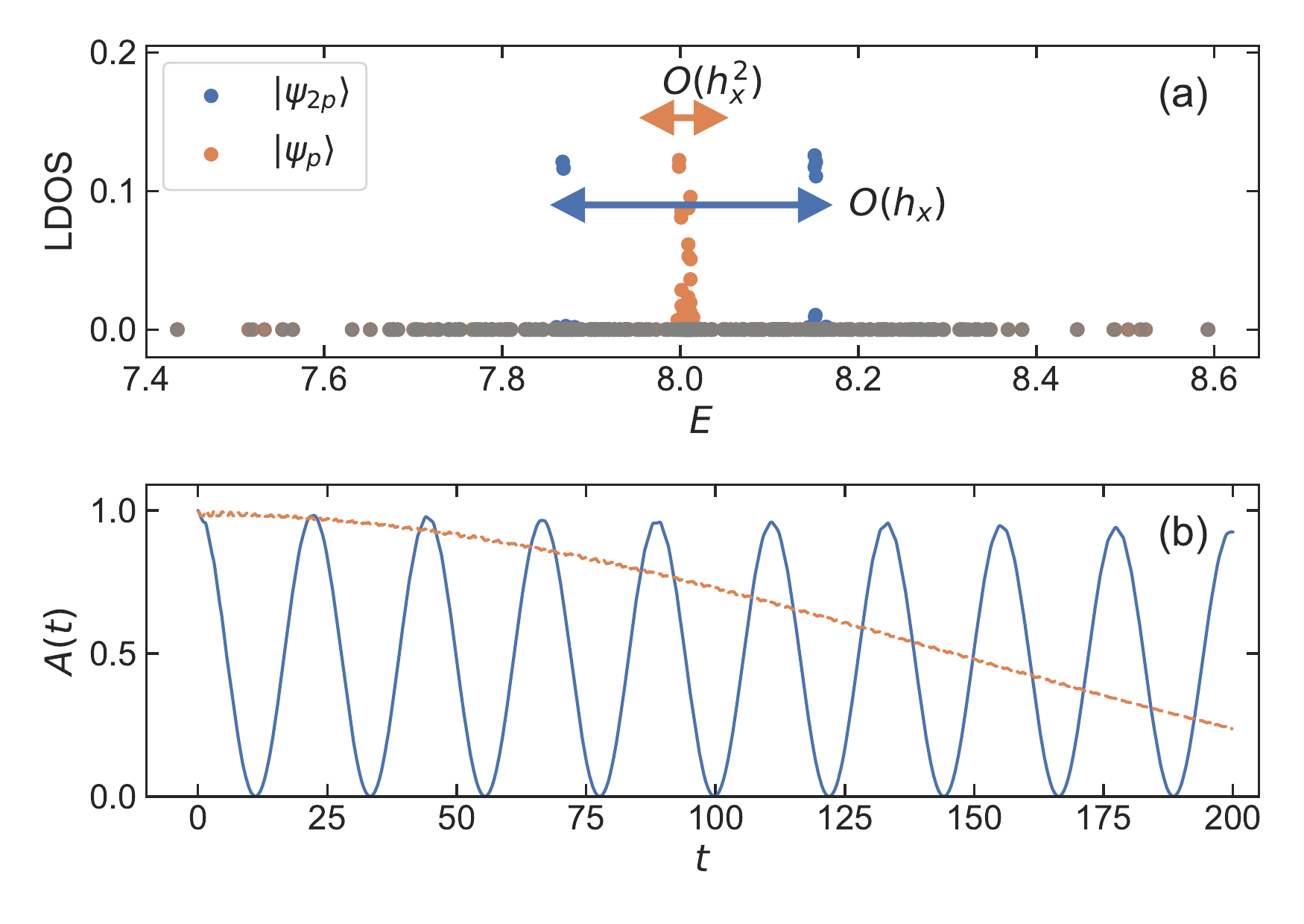}
	\caption{(a) LDOS and (b) autocorrelation for the single peak state $\left|\psi_p\right\rangle$, Eq.~(\ref{peak}), (dashed, light, orange) and a twin peaks state $\left|\psi_{2p}\right\rangle$, Eq.~(\ref{twinpeak}), (solid, dark, blue) with $L=8$ and $h_x=0.1$.}
	\label{fig:peaks_combined_arrows}
\end{figure}

In the small $h_x$ limit, any connections between basis states with a different potential will be strongly suppressed as $O(h_x^2)$. We can thus identify local spin configurations such that flipping a given spin does not change the potential, which is equivalent to requiring that flipping the spin does not alter the number of (un)equal neighbours (we recall we use periodic boundary conditions in both directions). This is achieved by configurations of the form
\begin{equation}\label{zeroflip}
\small{ \left| \begin{array}{ccccc} \cdots & n & \underline{n_a} & n & \cdots\\  \cdots & \cdot & 1-n & \cdot & \cdots \end{array} \right\rangle},
\end{equation}
for both $n=0,1$. Here, the underline denotes the spin which can be flipped ($n_a\rightarrow 1-n_a$). 
The $\left| \mathbb{Z}_2\right\rangle$ states are special, in that every site of them is as in Eq.~(\ref{zeroflip}). We will now consider the other extreme, where (almost) no site has this local configuration. Firstly, there are basis states that do not have this pattern anywhere, for example
\begin{equation}\label{peak}
\left| \psi_p\right\rangle = \sum_{i,j} T_x^i\, T_y^j \small{ \left| \begin{array}{cccccccc} 1 & 1 & 1 & 1 & 0 & 0 & 1 & 0 \\ 1 & 1 & 1 & 0 & 0 & 0 & 0 & 1 \end{array} \right\rangle},
\end{equation}
with $V = 8$. The action of $H_x$ on any site of this basis state will change the potential, so up to first order in $h_x$ the energy will not be changed and consequently the LDOS$_{\left|\psi_p \right\rangle}$ is sharply peaked around $E=V+ O(h_x^2)$ for small values of $h_x$, see Fig.~\ref{fig:peaks_combined_arrows}a. This will be called a peak state. For large $L$ the number of peak states scales as $\phi^{2L}$, where $\phi=(1+\sqrt{5})/2\approx 1.618$ is the golden ratio~\cite{Suppl}.

Secondly, consider a basis state where the configuration~(\ref{zeroflip}) only occurs once, such as
\begin{equation}\label{trippeak}
\left| \psi_{3p}\right\rangle = \sum_{i,j} T_x^i T_y^j \small{ \left| \begin{array}{cccccccc}  1 & 1 & 1 & 0 & 1 & 1 & 0 & 1\\ 1 & 1 & 0 & 1 & 1 & 1 & \underline{1} & 1\end{array} \right\rangle}.
\end{equation}
Flipping the marked site results in the basis state
\begin{equation}\label{twinpeak}
\left| \psi_{2p}\right\rangle = \sum_{i,j} T_x^i T_y^j \small{ \left| \begin{array}{cccccccc}  1 & 1 & 1 & 0 & 1 & 1 & \underline{0} & 1\\ 1 & 1 & 0 & 1 & 1 & 1 & \underline{0} & 1\end{array} \right\rangle},
\end{equation}
which has two sites that could be flipped without a change in the potential. Flipping the lower site returns $\ket{\psi_{2p}}$ to state~(\ref{trippeak}) while flipping the other one leads to
\begin{equation}\label{trippeak2}
\left| \psi_{3p'}\right\rangle = \sum_{i,j} T_x^i T_y^j \small{ \left| \begin{array}{cccccccc}  1 & 1 & 1 & 0 & 1 & 1 & \underline{1} & 1\\ 1 & 1 & 0 & 1 & 1 & 1 & 0 & 1\end{array} \right\rangle}.
\end{equation}
Here, the energy conserving flip connects again to $\left| \psi_{2p}\right\rangle$, such that there is an effective subspace of the states $\{\left|\psi_{3p} \right\rangle \leftrightarrow \left|\psi_{2p} \right\rangle \leftrightarrow \left|\psi_{3p'} \right\rangle\}$ connected by $H_x$. The eigenvalues are easily found to be, to leading order in $h_x$, $V$ and $V\pm\sqrt{2}\,h_x$, where $V$ is the potential of all three basis states, which in this example is $V=8$. The eigenvector with energy $V$ has no overlap with the state $\left| \psi_{2p}\right\rangle$, resulting in LDOS$_{\left| \psi_{2p}\right\rangle}$ featuring only two peaks at $E=V\pm\sqrt{2}h_x+ O(h_x^2)$, see Fig.~\ref{fig:peaks_combined_arrows}a. The other two basis states $\left|\psi_{3p} \right\rangle$ and $\left|\psi_{3p'} \right\rangle$ do have an overlap with the $E=V$ eigenvector and therefore their LDOS consist of three peaks, at $E=V\pm\sqrt{2}h_x+ O(h_x^2)$ and $V+ O(h_x^2)$. These special overlaps have consequences for the time evolution after a quench from one of these basis states. On one hand, for the single peak the autocorrelation~(\ref{autocor}) will decay slowly, on a time scale $T \propto 1/O(h_x^2)$ corresponding to the width of the peak. On the other hand, the twin peak state results in a clear oscillation of the autocorrelation with period given by the energy separation between the two peaks, $T\approx 2\pi/(2\sqrt{2}h_x)$, see Fig.~\ref{fig:peaks_combined_arrows}b. When increasing $h_x$, the peaks in the LDOS will broaden and consequently $\langle A\rangle$ will decrease. In this sense, the ETH-breaking behaviour of the peak states is not robust with respect to the increase of $h_x$, in contrast to the $\left| \mathbb{Z}_2\right\rangle$ states.

Importantly, exploiting the constraint Eq. (\ref{zeroflip}) leading to Eqs. (\ref{peak})-(\ref{trippeak2}) can be generalized to the whole 2D plane, as well as to higher dimensions. As an explicit example, lets consider the state
\begin{equation}\label{peakstate4x4}
\ket{\psi_p^{4 \times 4}}=
\left|\footnotesize{\begin{array}{cccc}
0 & 1 & 0 & 1 \\ 
0 & 0 & 1 & 1 \\ 
0 & 1 & 0 & 1 \\ 
1 & 1 & 0 & 0
\end{array}} \right\rangle.
\end{equation}
We observe that there is no site that is connected to two equal and two unequal neighbours. Therefore flipping any single spin necessarily results in the change of the energy and, in analogy to (\ref{peak}), we conclude that Eq.~(\ref{peakstate4x4}) is a peak state. Next, we note that repeating (\ref{peakstate4x4}) in $x,y$ directions, such that the system size becomes $4m\times4n$, $m,n \in {\mathbb N}$, results again in a peak state. This construction can be extended to a hypercubic lattice in higher dimensions. One systematic way to accomplish this is by recursively creating the $d+1$ dimensional lattice by layering $4r$ copies, with $r \in \mathbb{N}$, of the $d$-dimensional lattice, with the $d=2$ ``seed'' lattice being for example the state $\ket{\psi_p^{4 \times 4}}$ extended to $4m\times4n$ sites. Denoting the $i$-th $d$-dimensional layer as $\mathcal{L}_i$, acting with the particle-hole conjugate operator $Q$ on every site of a $\mathcal{L}_i$ with $i \mod 4 = 2$ or $i \mod 4 = 3$ leads to every layer being connected to one layer with all inter-layer neighbours equal and one with all inter-layer neighbours unequal, such that for every site the total number of equal and unequal neighbours is still different and therefore this is a peak state.


\textit{Outlook.}---In this work we have analyzed transverse field Ising ladder and identified families of initial product states resulting in ETH-violating behaviour which can be interpreted as quantum many-body scars. The present analysis allowed to identify scarred initial states in higher dimensions on a square lattice. This opens a way for generalizations to other lattice geometries, for example the honeycomb lattice featuring frustrated ground states~\cite{Coletta_2011_PRB}. It would be also interesting to consider the action of the longitudinal field. In one dimension this results in the meson excitations~\cite{James_2019_PRL, Robinson_2019_PRB}, which are low-variance states~\cite{Robinson_private} akin to the $\ket{{\mathbb Z}_2}$ and peak states hinting to their direct relation to many-body scars. While tackling the outlined questions is theoretically challenging, they can be readily addressed with Rydberg atom based platforms and present thus an ideal test-bed for quantum simulations beyond simple chain geometries.

We are very grateful to Neil Robinson, Ward Vleeshouwers and Vladimir Gritsev for stimulating discussions. This work is part of the Delta ITP consortium, a program of the Netherlands Organisation for Scientific Research (NWO) that is funded by the Dutch Ministry of Education, Culture and Science (OCW).

\bibliographystyle{apsrev4-1}
\bibliography{scars_ising_2D_arxiv}

\clearpage

\onecolumngrid
\appendix

\section*{Supplementary Material}

\section{Degeneracy of the eigenstates}

\subsection{Direct counting}

\begin{figure}
		\centering
		\includegraphics[width=0.2\textwidth]{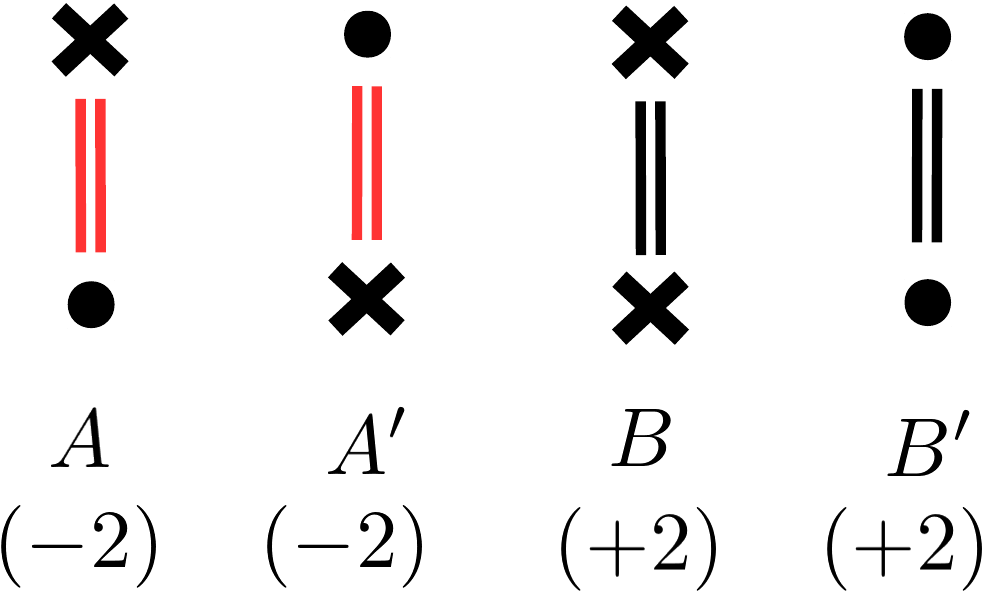}
		\caption{Possible configurations of the rungs of the ladder with the respective energies indicated in parenthesis. The black dot (cross) stands for 0 (1) spin, the double lines are due to the periodic boundary conditions and their colour denote the contribution to the energy - each red (black) line contributes energy -1 (+1).}	
		\label{fig:notation}
	\end{figure}

	In this section we evaluate the degeneracies of the eigenstates of the Ising ladder without the transverse field.
	Consider the Hamiltonian
	\begin{equation}
	H =  H_z =\sum_{\left\langle i,j\right\rangle}\sigma^z_i \sigma^z_j
	\end{equation}
	acting on the basis states (\ref{basis_definition}) given in terms of local occupation numbers ($\left| 0\right\rangle$ and $\left| 1\right\rangle$) on each site.
	As noted in the main text, two neighbouring sites with equal (unequal) occupation number will contribute energy $+1$ ($-1$). It is convenient to introduce the following symbols for the four possible configurations of a rung, see Fig. \ref{fig:notation}, namely
	\begin{equation}
	A=\begin{pmatrix} 1 \\ 0 \end{pmatrix}, \; A'=\begin{pmatrix} 0 \\ 1 \end{pmatrix}, \; B=\begin{pmatrix} 1 \\ 1 \end{pmatrix}, \; B'=\begin{pmatrix} 0 \\ 0 \end{pmatrix}
	\end{equation}
	with respective energies -2,-2,+2,+2 due to periodic boundary conditions (PBC). Each basis state thus corresponds to a string formed by letters from the set $\{A,A',B,B'\}$ so that we have reduced the problem of counting the degeneracies of the state of a given energy to enumerating the number of strings yielding that energy subject to the following rules. We denote the set $\{A,A'\}$ as the $A-$subset and $\{B,B'\}$ as the $B-$subset. On top of the energies associated with the rungs, there are also energies associated to the links connecting two letters. When two letters are the same, the link contributes $E=+2$ (e.g. $A-A$), two different letters of the same subset contribute $E=-2$ (e.g. $A-A'$) and links between letters from different subsets contribute $E=0$ (e.g. $A-B$).

The extremal energies take values $E=\pm 4L$ and the spectrum is symmetric. Focusing on $E=4L$ states, they are two-fold degenerate and, as stated in the main text, they correspond to all sites occupied by either 0 or 1 spins. Using the letter notation, they correspond to $B-B-\ldots-B-$ and $B'-B'-\ldots-B'-$ respectively. Here, the last link represents the link connecting the last and first site due to the PBC. Each basis state thus corresponds to a one-dimensional graph, where both the vertices (the letters) and the edges (the links) carry energy. This motivates us to introduce a symbolic two-line notation, where the lower line describes the energy of the vertices and the upper line that of the edges. For instance, considering $L=4$ we have for the $E=4L$ states
	\begin{equation}
	B-B-B-B- \; \cong \; B'-B'-B'-B'- \; \cong \;
	\begin{pmatrix}
	+_{12} & +_{23} & +_{34} & +_{41} \\
	+_{1} & +_{2} & +_{3} & +_{4}   
	\end{pmatrix},
	\label{eq:mnotation}
	\end{equation}
	where the indices $i$ in $+_{i}$ label the sites and $+_{ij}$ denotes the energy of the link connecting sites $i$ and $j$. In what follows we drop these indices for simplicity and just use the symbols $\{+,-,0\}$ to denote an energy contribution of respectively $\{+2,-2,0\}$. We note that the assignment in Eq.~(\ref{eq:mnotation}) is not unique and the matrix notation represents an equivalence class of graphs, whose energy pattern is identical up to the translations $(T_x)^k$. The notation (\ref{eq:mnotation}) is useful mainly for bookkeeping purposes. To demonstrate this, we next evaluate the degeneracies of the first de-excited state of $E=4L-8$.
	
There are two different ways in which a state with $E=4L-8$ can be created starting from Eq.~(\ref{eq:mnotation}). Firstly, replacing one $B$ by an $A$ and secondly by changing a string of $l$ consecutive $B$-letters by a string of its complement in the $B$-subset (i.e. $B \rightarrow B'$ or $B' \rightarrow B$ within the string of length $l$). Taking $L=4$ as an example, the states with $E=4L-8$ are
	\begin{equation}
	\begin{pmatrix}
	0 & + & + & 0 \\
	- & + & + & +   
	\end{pmatrix}  {}\;\; {\rm deg}=16, \qquad 
	\begin{pmatrix}
	- & - & + & + \\
	+ & + & + & +   
	\end{pmatrix}  {}\;\; {\rm deg}=8,  \qquad
	\begin{pmatrix}
	- & + & - & + \\
	+ & + & + & +   
	\end{pmatrix}  {}\;\; {\rm deg}=4,
	\end{equation}
	where the numbers denote the respective degeneracies. For instance, the first class has $L$-fold degeneracy due to translation and there are two possibilities to choose the $A$-letter and two possibilities for the string of the $+$ vertices, i.e. $L\times 2 \times 2 = 4L$. Similar considerations for the two remaining classes lead to a general formula for the degeneracies of the first (de-)excited state
	\begin{equation}
	\label{eq:deg_1st}
	\begin{aligned}
	{\rm deg}_{E=4L-8} &= {\rm deg}_{E=-4L+8} \\
	&= 2 \left[ 2 L + L\left( \frac{L}{2}-1 \right) + \frac{L}{2} \right] \\
	&= L^2 + 3L.
	\end{aligned}
	\end{equation}

\subsection{Generating function}

In principle one can extend the above described direct counting to other energy manifolds, but the method becomes quickly cumbersome with increasing system size. Here we describe a recursive construction of a generating function (partition sum)
\beq
	Z(x) = \sum_n a_n x^n,
\eeq
where the coefficients $a_n$ are the degeneracies of a manifold of energy $n$. Lets first consider an \emph{open} chain which starts with a letter $A$ of energy -2. We thus attribute an element $x^{-2}$ to this configuration. One can now add one of the four possible letters to the right of $A$: $A$ (0), $A'$ (-4), $B$ (+2), $B'$ (+2) and similarly for having either $A'$, $B$ or $B'$ at the beginning of the chain. Here the numbers in the brackets denote the energy contribution from adding the given letter, i.e. the power of $x$. This method can be continued recursively by the following prescription
\begin{subequations}
\label{eq:Zalphabeta}
	\begin{align}
		Z_{A\alpha}^{(L+1)} &=\left(1 + x^{-4}\right) Z_{A\alpha}^{(L)} + 2 x^{-2} Z_{A \beta}^{(L)} \\
		Z_{A\beta}^{(L+1)} &= 2 x^2 Z_{A\alpha}^{(L)} + \left(1 + x^4\right) Z_{A \beta}^{(L)},
	\end{align}
\end{subequations}
with the seed $(Z_{A\alpha},Z_{A\beta})=(x^{-2},0)$. Here $Z_{A\alpha} = Z_{AA}+Z_{AA'}$, $Z_{A\beta}=Z_{AB}+Z_{AB'}$ and $Z_{A\mu}^{(L)}$ is a generating function of order $L$ in the recursion corresponding to adding a letter $\mu=A,A',B,B'$ to the right of the chain. Considering only $A$-letters, expressions analogous to Eqs. (\ref{eq:Zalphabeta}) can be obtained for the respective generating functions
\begin{subequations}
	\begin{align}
		Z_{AA}^{(L+1)} &= Z_{AA}^{(L)} + x^{-4} Z_{AA'}^{(L)} \\
		Z_{AA'}^{(L+1)} &= x^{-4} Z_{AA}^{(L)} + Z_{AA'}^{(L)}	
	\end{align}
\end{subequations}
	For later convenience we define
\beq
	\label{eq:delta_Z}
	\delta Z_{A\alpha}^{(L)} \equiv Z_{AA}^{(L)} - Z_{AA'}^{(L)} = x^{-2}(1-x^{-4})^{L-1}. 
\eeq	
The periodic boundary condition is then implemented by the mappings 
\beq
	\label{eq:Z_PBC}
	Z_{AA} \rightarrow x^2 Z_{AA},\; Z_{AA'} \rightarrow x^{-2} Z_{AA'}, \; Z_{AB} \rightarrow Z_{AB}, \; Z_{AB'} \rightarrow Z_{AB'}.
\eeq
Analogous mappings hold for $A'$ instead of $A$ at the beginning of the chain. Starting with $B,B'$, analogous mapping to (\ref{eq:Z_PBC}) holds again, with $x \rightarrow x^{-1}$, provided $L$ is \emph{even}, which is the case of our interest. This results in the expression for the generating function of a chain of length $L$
\beqa
\label{eq:Zeven}
	Z^{(L)} &=& 2 \left( f(x) + f(x^{-1})\right) \nonumber \\
	f(x) &=& x^2 Z_{AA}^{(L)} + x^{-2} Z_{AA'}^{(L)} + Z_{A\beta}^{(L)},
\eeqa
which can be readily evaluated with the help of the relations (\ref{eq:Zalphabeta}) and (\ref{eq:delta_Z}). This result can be however further manipulated as follows. First we note, that casting the relation (\ref{eq:Zalphabeta}) in the matrix form, it can be diagonalized as
\beq
	\vec{Z}^{(L+1)}=M \vec{Z}^{(L)} = 
	\begin{pmatrix}
		1 + x^{-4} & 2 x^{-2} \\
		2 x^2 & 1 + x^4
	\end{pmatrix}
	\vec{Z}^{(L)} \, \rightarrow \, v_\pm^{(L+1)} = \lambda_\pm v_\pm^{(L)},
\eeq
where $\vec{Z}^{(L)} = (Z_{A\alpha}^{(L)},Z_{A\beta}^{(L)})^{\rm T}$,
\beq
	\lambda_\pm = 2 x^{-4} \left( 1+2 x^4 + x^8 \pm s \right), \; s=\sqrt{1+14 x^8+x^{16}}
\eeq
are the eigenvalues of $M$ and $v_\pm$ the corresponding eigenvectors. Expressing $Z_{A\alpha}, Z_{A \beta}$ in terms of $v_{\pm}$, substituting to (\ref{eq:Zeven}) and using again (\ref{eq:delta_Z}), after some algebraic manipulations we obtain a closed-form expression for the generating function
\beq
	Z^{(L)} = \frac{s}{2}\left( \lambda_+^{L-1} - \lambda_-^{L-1} \right) + \frac{1}{2} \left(x^{-4} + 2 + x^4 \right) \left( \lambda_+^{L-1} + \lambda_-^{L-1} \right) + \left(x^{-2L} + x^{2L} \right) \left(x^{2} - x^{-2} \right)^L.
	\label{eq:Zclosedform}
\eeq
Applying first (\ref{eq:Zclosedform}) to the example of the ladder of $L=4$ we find
\beq
	Z^{(L=4)}=2 \left(x^{-16}+14 x^{-8}+24 x^{-4} + 50 +24 x^4 + 14 x^8+ x^{16}\right),
\eeq
which gives ${\rm deg}_{E=\pm 4L}=2$ and ${\rm deg}_{E=\pm(4L-8)}=28$ as it should, cf. Eq. (\ref{eq:deg_1st}). Furthermore, we can now evaluate the degeneracy of the $E=0$ manifold, which gives ${\rm deg}_{E=0}=100$ and for $L=6,8$ studied in this work we get ${\rm deg}_{E=0}=1188,15876$.

\section{Autocorrelation of the (a)typical states}
\begin{figure}[h]
	\centering
	\includegraphics[width=.45\linewidth]{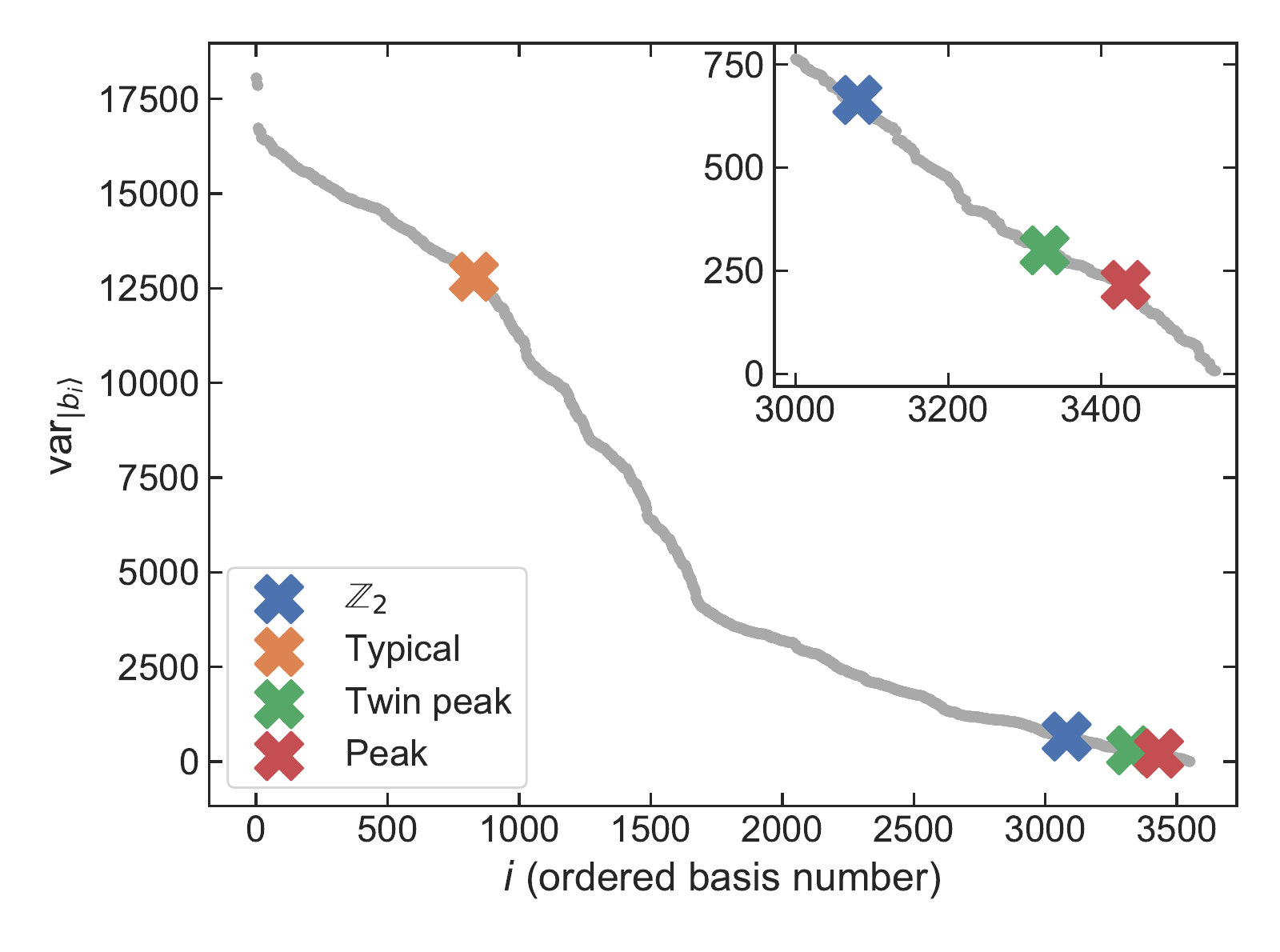}
	\caption{The weight variances for all basis states with $V=0, \pm4, \pm8$ of the $L=8$ and $h_x=0.1$ system. The four specific states considered in the text are denoted by a cross. The inset shows the lowest variance states.}	
	\label{fig:var_weights}
\end{figure}

The difference between a typical state and an atypical $\left| \mathbb{Z}_2\right\rangle$ state results in revivals of the autocorrelation $A_{\left| \psi_0\right\rangle}(t) = |\left\langle \psi_0|e^{-iHt}|\psi_0 \right\rangle|^2$, Eq.~(\ref{autocor}), for the latter in contrast to the former. Writing the initial state as a superposition of the eigenstates $\left| \psi_0\right\rangle = \sum_j c_j \left| v_j\right\rangle$, the autocorrelation can be expressed as
\begin{equation}\label{autocor_appendix}
\begin{aligned}
A(t)_{\left| \psi_0\right\rangle} 
= \sum_j |c_j|^4 + 2 \sum_{j<k}|c_j|^2 |c_k|^2 \cos\left(E_{jk}t\right),
\end{aligned}
\end{equation}
where $E_{jk} = E_j-E_k$ is the energy difference between eigenenergies of the eigenstates $v_j,v_k$. The properties of the autocorrelation thus immediately follow from the weights $w_j=|c_j|^2$ as well as the energy differences $E_{jk}$ between the states with non-zero weights. 
First we analyze the distribution of the weights for each basis state $\ket{\psi_0}=\ket{b_i},\, \forall i$ by ordering them from high to low values such that $w_j \geq w_{j+1}$. We can then quantify the corresponding distribution of the weights using the variance
\beq
 {\rm var}_{\ket{b_i}} = \sum_j(j-\bar{n})^2 w_j^{(i)} \text{ with } \bar{n} = \sum_j j w_j^{(i)},
  \label{eq:var}
\eeq
where $\ket{b_i}=\sum_j c_j^{(i)} \ket{v_j}$. In Fig.~\ref{fig:var_weights} we show the ordered variances for $L=8$ and for basis states with potential $|V|=|\braket{b_i | H_z | b_i}| \leq 8$ as we are interested in the mid-spectrum states.
As can be seen, the $\mathbb{Z}_2$ state and the typical state~(\ref{psityp}) that were used in Fig.~\ref{fig:time_evolution_Z2_other_L8_combinedhx}, have respectively a low and high variance. The basis states with the lowest variance belong to the (twin) peak states, including the ones used in the main text [Eq. (\ref{peak}) and (\ref{trippeak})].
We note that the low variance of the weights alone is not sufficient to guarantee quasiperiodic revivals of the autocorrelation (so that not all low-variance states would be classified as QMBS) and has to be supplemented by the quasi equidistant energy spacings between the eigenstates with the highest weight as in the case of the $\ket{{\mathbb Z}_2}$ state.

\subsection{Autocorrelation vs. $L$ and $h_x$}
\begin{figure}[h]
	\centering
	\includegraphics[width=.9\linewidth]{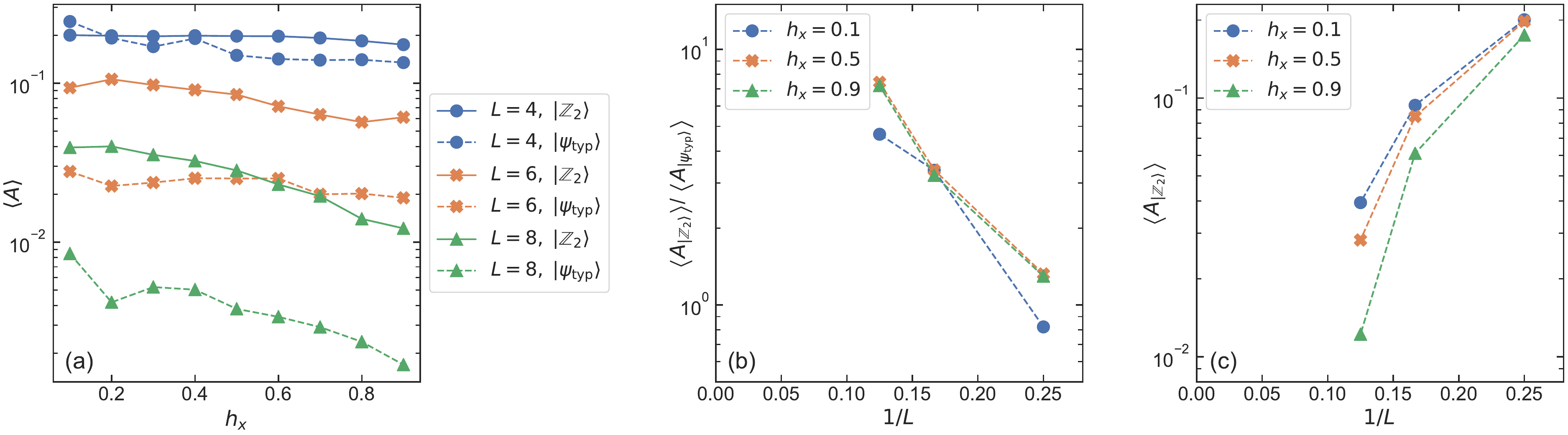}
	\caption{(a) The average autocorrelation for a $\left| \mathbb{Z}_2\right\rangle$ state (solid line) and the typical state $\ket{\psi_{\rm typ}}$ Eq.~(\ref{psityp}) (dotted line) for $L=4,6,8$ (blue dots, orange crosses, green triangles) as a function of $h_x$. (b) The ratio $\left\langle A_{\left| \mathbb{Z}_2\right\rangle} \right\rangle / \left\langle A_{\ket{\psi_{\rm typ}}} \right\rangle$ and (c) $\left\langle A_{\ket{\mathbb{Z}_2}} \right\rangle$ as a function of $1/L$ for $h_x=0.1, 0.5, 0.9$. In all three subplots the same data is used with $t_0=133$ and $\tau=500$.}
	\label{fig:avgautocorcombined}
\end{figure}

Here we study numerically the dependence of the average autocorrelation $\langle A \rangle=\lim_{\tau\rightarrow \infty} \tau^{-1}\int_{t_0}^\tau A(t)\,dt$ on the system size $L$ and the transverse field $h_x$, where we choose $t_0>0$ such that the initial decay of $A(t)$ is not included in the average and $\tau$ large enough that $\langle A \rangle$ becomes $\tau$-independent. In Fig.~\ref{fig:avgautocorcombined}a we show $\braket{A}$ vs. $h_x$ for $L=4,6,8$ (blue, orange, green) for both an atypical ($\ket{{\mathbb{Z}_2}}$, solid lines) and a typical ($\ket{\psi_{\rm typ}}$, Eq.~(\ref{psityp}), dashed lines) initial state. We see the decrease of the average autocorrelation with $h_x$ for all system sizes, which is compatible with the decrease of the fidelity of the SMA states, describing idealized scar behaviour, with respect to the system eigenstates, cf. Fig.~\ref{fig:overlapSMA_all}. 

Comparing $\braket{A}$ for the $\ket{{\mathbb Z}_2}$ and the typical state, we see that $\braket{A_{\ket{{\mathbb Z}_2}}} > \braket{A_{\ket{\psi_{\rm typ}}}}$ for all values of $h_x$ and $L$ (except for $L=4$ and $h_x=0.1$, which is likely a finite-size effect). In Fig.~\ref{fig:avgautocorcombined}b we plot $\braket{A_{\ket{{\mathbb Z}_2}}}/\braket{A_{\ket{\psi_{\rm typ}}}}$ vs. $1/L$ for $h_x=0.1, 0.5, 0.9$ and see a clear increase of the ratio with the system size. We have verified that the higher value of $\braket{A_{\ket{{\mathbb Z}_2}}}$ can be indeed attributed to the revivals of $A_{\ket{{\mathbb Z}_2}}(t)$. At the same time, as can be seen in Fig.~\ref{fig:avgautocorcombined}c, $\left\langle A_{\left| \mathbb{Z}_2\right\rangle}\right\rangle$ decreases with $L$, similar to results on the PXP model~\cite{Lin2018}. In conclusion, while the quantitative details depend on the choice of the typical state, the numerical simulations suggest that the atypical behaviour of the $\left| \mathbb{Z}_2\right\rangle$ states is robust in the sense that the ratio $\left\langle A_{\left| \mathbb{Z}_2\right\rangle} \right\rangle / \left\langle A_{\text{typ}} \right\rangle \gg 1$ for all $h_x$ and system sizes $L$ and in fact increases with system size. Further studies, including e.g. the use of SMA construction or Mazur inequalities~\cite{Mazur1969,Suzuki_1971,Caux_2011_JstatPhys,Sala_2020_PRX} are required to assess the nature of the autocorrelations in the thermodynamic limit.

\section{Proof of zero energy state}\label{proofzero:a}
To show that $\left| \psi_{E=0}'\right\rangle$, Eq.~(\ref{exactzero}), is a zero energy eigenvector independent of $h_x$, i.e. $H\left| \psi_{E=0}'\right\rangle=0$, we first consider the action of $H_z$ by expanding it as $H_z = \sum_i H^z_{i}$ with
\begin{equation}\label{Hzlocal}
H_i^z \equiv \sigma^z_{i,1}\,\sigma^z_{i+1,1} + 2\,\sigma^z_{i+1,0}\,\sigma^z_{i+1,1} + \sigma^z_{i+1,0}\,\sigma^z_{i+2,0}.
\end{equation}
$H^z_i$ is a local operator that only acts on the sites occupied by two neighbouring singlets $s_i$ and $s_{i+1}$ such that
\begin{equation}\label{Hzsisi+1}
\begin{aligned}
H_z \left| \psi_{E=0}'\right\rangle  &= \sum_i H^z_i \left| \psi_{E=0}'\right\rangle 
= \sum_i \left(\prod_{j\neq i,i+1} s_j\right) H_i^z s_i s_{i+1} \left|\emptyset \right\rangle .
\end{aligned}
\end{equation}
The product of the two singlets written out in basis states is
\begin{equation}\label{singlet_neigh}
\begin{aligned}
s_{i} s_{i+1}\left|\emptyset\right\rangle &= \left(c^\dagger_{i,1} - c^\dagger_{i+1,0}\right)\left(c^\dagger_{i+1,1} - c^\dagger_{i+2,0}\right)\left|\emptyset\right\rangle\\
&=  \footnotesize \left| \begin{array}{ccc}  1 & 1 & \cdot \\    \cdot & 0 & 0 \end{array} \right\rangle - \left| \begin{array}{ccc} 1 & 0 & \cdot \\ \cdot & 0 & 1 \end{array} \right\rangle - \left| \begin{array}{ccc} 0 & 1 & \cdot \\ \cdot & 1 & 0 \end{array} \right\rangle + \left| \begin{array}{ccc} 0 & 0 & \cdot \\ \cdot & 1 & 1 \end{array} \right\rangle,
\end{aligned}
\end{equation}
where in the last line only the occupation numbers of the relevant positions (i.e. those occupied by the singlets $s_i$ and $s_{i+1}$) are denoted. In each of these basis states there are two neighbouring pairs with the same and two pairs with a different occupation number, such that the action of $H^z_i$ on all these basis states is zero: $H^z_i s_i s_{i+1} \left| \emptyset\right\rangle = 0$. Entering this result in Eq. (\ref{Hzsisi+1}) then shows that $H_z \left| \psi_{E=0}'\right\rangle=0$.
Secondly, consider the action of $H_x$ in a similar way by using the expansion $H_x = \sum_i H^x_i$ with $H_i^x = \sigma^x_{i,1} + \sigma^x_{i+1,0}$ that only acts on the sites of a single singlet. Then
\begin{align}\label{H_x_singlet}
H_x \left| \psi_{E=0}'\right\rangle &= h_x \sum_{i} \left(\prod_{j\neq i} s_j\right) H^x_i s_i \left| \emptyset\right\rangle
\end{align}
and
\begin{equation}\label{actionHxi}
\begin{aligned}
H^x_i s_i \left| \emptyset\right\rangle &= (\sigma^x_{i,1} + \sigma^x_{i+1,0})(c^\dagger_{i,1} - c^\dagger_{i+1,0})\left| \emptyset\right\rangle \\
&=  (1-1+c^\dagger_{i,1}c^\dagger_{i+1,0}-c^\dagger_{i,1} c^\dagger_{i+1,0})\left| \emptyset\right\rangle \\
&= 0.
\end{aligned}
\end{equation}
Because of Eq. (\ref{actionHxi}) it follows that $H_x \left| \psi_{E=0}'\right\rangle=0$ and therefore $H\left| \psi_{E=0}'\right\rangle=0$. This concludes the proof that the state (\ref{exactzero}) is indeed a zero energy state, independent of $h_x$.\\

\subsection{Other $h_x$-independent eigenstates}
While the $\ket{\psi'_{E=0}}$ state is an exact eigenstate of the Hamiltonian (\ref{Hamiltonian}), we could identify other eigenstates, which are independent of the transverse field even for $E \neq 0$. For example, for $L=4$, non-degenerate eigenstates exist with energy $E=\pm4$, which are explicitly
\begin{align}\label{psi-4}
\left| \psi_{\footnotesize E=-4}\right\rangle &= \sum_{i,j,r,q} T_x^{i} T_y^{j} R^r Q^q (-1)^r  \small{ \left| \begin{array}{cccc} 1 & 1 & 1 & 0 \\ 1 & 0 & 0 &1 \end{array} \right\rangle},\\
\left| \psi_{\footnotesize E=4}\right\rangle &= \sum_{i,j,r,q} T_x^{i} T_y^{j} R^r Q^q (-1)^r  \small{ \left| \begin{array}{cccc} 1 & 1 & 0 & 0 \\ 1 & 0 & 0 &0 \end{array} \right\rangle},
\end{align}
with $T_x$ and $T_y$ the translation operator, $R$ the reflection operator ($c^\dagger_{x,y} \rightarrow c^\dagger_{L-x,y}$) and $Q$ the particle-hole inversion operator ($n_{x,y} \rightarrow 1-n_{x,y}$), with $r\in\{0,1\}$ and $q\in\{0,1\}$. Additionally, we could identify $h_x$-independent $E=\pm4$ ($E=-4$) eigenstates for a $L=3$ ladder ($L_x=3,L_y=4$ system). However, analogous $E=\pm4$ eigenstates do not seem to exist for larger system sizes. Moreover, we do not consider these eigenstates any further, because in contrast to the $\left|\psi_{E=0} \right\rangle$ state, single mode approximation excitations on top of these states do not lead to non-thermalizing behaviour.

\section{Systematics for computing energies of the SMA states}

In this section we apply the forward scattering approximation to systematically construct approximate eigenstates $\left| w_{(j)}\right\rangle$, with $j$ denoting the order of the approximation, to the exact eigenstates $\left| v'_{\text{\scriptsize{QMBS}},n}\right\rangle$ of $H'$, Eq.~(\ref{Hamiltonian_PXP}),
$
H' = \sum_{\langle i,j \rangle} \sigma^z_{i} \sigma^z_{j} + h_x \sum_i P\sigma^x_{i}P.
$ 
We are interested on the action of $H'$ on the space that is projected onto the states with $V=0$ in order to analyze the energy difference between the SMA states on this space. For simplicity we will set $h_x=1$, such that $H'\rightarrow \sum_i P\sigma_i^x P$.  It is useful to rewrite the basis states in terms of the building blocks
\begin{equation}
\begin{aligned}
s^- = \small{\left| \begin{array}{cc} 0 & \cdot \\ \cdot & 1 \end{array} \right\rangle- \left| \begin{array}{cc} 1 & \cdot \\ \cdot & 0 \end{array} \right\rangle}, \quad
&
s^+ = \small{ \left| \begin{array}{cc} 0 & \cdot \\ \cdot & 1 \end{array} \right\rangle + \left| \begin{array}{cc} 1 & \cdot \\ \cdot & 0 \end{array} \right\rangle} ,
\\\
t^- =  \small{\left| \begin{array}{cc} 0 & \cdot \\ \cdot & 0 \end{array} \right\rangle - \left| \begin{array}{cc} 1 & \cdot \\ \cdot & 1 \end{array} \right\rangle}, \quad
& 
t^+ =\small{ \left| \begin{array}{cc}0 & \cdot \\ \cdot & 0 \end{array} \right\rangle + \left| \begin{array}{cc} 1 & \cdot \\ \cdot & 1 \end{array} \right\rangle, }
\end{aligned}
\end{equation}
where the normalization constant is omitted for clarity. The state $\left| \psi'_{\text{SMA},1}\right\rangle$~(\ref{SMAn}) is built up starting from the reference state $\left| \psi_{E=0}'\right\rangle$~(\ref{exactzero}), which is 
\begin{equation}\label{vacuum_kj}
\left| \psi_{E=0}'\right\rangle  = \left[ \ldots s^- s^- \ldots \right].
\end{equation}
Acting with the $O^\pm$ operators on $\left| \psi_{E=0}'\right\rangle $ in a translationally invariant way produces the first SMA states
\begin{align}
\label{eq:psiprime_SMA}
\left| \psi'_{\text{SMA},\pm1}\right\rangle =  &{1 \over \sqrt{2}} P \left[ \ldots s^- s^+ s^+ s^- \ldots \right]_{k_x=0} 
\ \pm\  {1 \over 2} P \left[\ldots s^- (s^+ t^+ + t^+ s^+) s^-\ldots\right]_{k_x=0}. 
\end{align}
This is independent of the length of the ladder $L$, such that the dots indicate that on all other sites is a $s^-$. Note that this state does not have $k_y=0$. In the remainder we will omit the explicit $k_x=0$ after each state, but all further states in this appendix will be with $k_x=0$. We can make use of the lemma
\begin{equation}
\begin{aligned}
&P\left[s^- t^+ s^+ \right] = {1 \over 2} \left[s^- t^+ s^+ + s^+ t^+ s^-\right] \text{\quad and\quad }P\left[s^- t^+ s^- \right] = {1 \over 2} \left[s^- t^+ s^- + s^+ t^+ s^+\right]
\end{aligned}
\end{equation}
to rewrite the state (\ref{eq:psiprime_SMA}) as
\begin{equation}
\begin{aligned}
\left| \psi'_{\text{SMA},\pm1}\right\rangle
=  &\pm {1 \over 4} \left[\ldots s^- (s^- s^+ t^+ s^- + s^- s^- t^+ s^+\right.
\left.+ \, s^- t^+ s^+ s^- + s^+ t^+ s^- s^-) s^- \ldots\right]  
+ {1 \over \sqrt{2}} \left[ \ldots s^- s^+ s^+ s^- \ldots \right] \\
= &\pm {1 \over 2} \left[\ldots s^- (s^+ t^+ + t^+ s^+) s^- \ldots\right]  + {1 \over \sqrt{2}} \left[ \ldots s^- s^+ s^+ s^- \ldots \right].
\end{aligned}
\end{equation}
Working out the terms results in 
\begin{equation}
\begin{aligned}
\left| \psi'_{\text{SMA},\pm1}\right\rangle
&= {\pm 1\over \sqrt{8}}\left[\ldots s^- \left( \left| \small{ \begin{array}{cccc} 1 & 0 & 1 & \cdot \\ \cdot & 0 & 0 & 0  \end{array}} \right\rangle\right.\right.
- \left|\small{\begin{array}{cccc} 0 & 1 & 0 & \cdot \\ \cdot & 1 & 1 & 1 \end{array} }\right\rangle  + \left| \small{\begin{array}{cccc} 1 & 1 & 1 & \cdot \\ \cdot & 0 & 1 & 0  \end{array} }\right\rangle 
- \left|\small{\begin{array}{cccc} 0 & 0 & 0 & \cdot \\ \cdot & 1 & 0 & 1 \end{array} }\right\rangle  \Bigg) s^- \ldots \Bigg]  \\
&\qquad+ {1 \over \sqrt{2}} \left[ \ldots s^- s^+ s^+ s^- \ldots \right] \\
& = {1 \over \sqrt{2}} (\pm \left|w_{st+ts}\right\rangle + \left|w_{ss} \right\rangle ).
\end{aligned}
\end{equation}
This wave function is an approximation to the exact eigenstates of the Hamiltonian. In order to get the energy of the excited state from the approximation, we will use the forward scattering approximation (FSA). We will take the state $\left| \psi'_{\text{SMA},1}\right\rangle$ as the initial vector of the FSA and which we call $\left|w_{(0)} \right\rangle$, where the index denotes the order of the FSA. In leading order, the energy of the excited state $\left|w_{(0)} \right\rangle$ is $E_{(0)}=\pm2\sqrt{2}=\pm2.828...$. We will now calculate the first corrections for the vectors and energies, which are independent of $L$.

\subsection{Finding $\left|w_{(1)} \right\rangle$}
As a first step we act on $\left|w_{(0)} \right\rangle$ with $H'$ and find $H' \left|w_{(0)} \right\rangle = 2\sqrt{2} \left|w_{(0)} \right\rangle + \left|w_{(1)} \right\rangle$, with
\begin{equation}
\begin{aligned}
\left|w_{(1)} \right\rangle & =  {1 \over 2}\left[ \ldots s^- \left(   \left| \small{\begin{array}{ccccc} 0 & 1 & 0 & 0 & \cdot \\ \cdot & 1& 1 & 0 & 1  \end{array}} \right\rangle \right. \right.
+ \left|\small{\begin{array}{ccccc} 1 & 0 & 1 & 1 & \cdot \\ \cdot& 0 & 0 & 1 & 0 \end{array}} \right\rangle
-\left|\small{ \begin{array}{ccccc} 1 & 1 & 1 & 0 & \cdot \\ \cdot & 0 & 1 & 1 & 1  \end{array}}\right\rangle
\left.\left.- \left|\small{\begin{array}{ccccc} 0 & 0 & 0 & 1 & \cdot\\ \cdot & 1 & 0 & 0 & 0 \end{array} }\right\rangle
\right) s^- \ldots \right]\\
&= \left|w_{tt} \right\rangle.
\end{aligned}
\end{equation}
The FSA matrix $M_{\rm FSA}$ with matrix elements $M_{i,j} = \left\langle w_{(i)}|H' |w_{(j)} \right\rangle$ becomes up to this order
\begin{equation}
M_{\rm FSA}^{(1)}= \left( \begin{array}{cc} 2\sqrt{2} & 1 \\ 1  & 0 \end{array} \right).
\end{equation}
It has as the largest eigenvalue $E_{(1)}=\sqrt{2}+\sqrt{3} = 3.146...$, which is the first-order approximation of the energy of the state $\ket{v'_{{\rm QMBS},1}}$.

\subsection{Finding $\left|w_{(2)} \right\rangle$}
We now apply $H'$ on $\left| w_{(1)}\right\rangle$ to obtain the next order correction. Acting on the inner indices of the blocks of four gives, after subtracting $\left|w_{(0)} \right\rangle$,
$$
H' \left|w_{(1)} \right\rangle - \left|w_{(0)} \right\rangle = \sqrt{3 \over 2} \left|w_{(0)}' \right\rangle - {1 \over \sqrt{2}} \left|w_{ss}  \right\rangle
$$
with  
\begin{equation}
\begin{aligned}
\left| w_{(0)}'\right\rangle &={1 \over 4\sqrt{3}}  
\Bigg[ \ldots s^- \Bigg(  3  \left|\small{\begin{array}{cccccc} 0 & 0 & 0 & 0 & 1 & \cdot \\ \cdot & 1 & 1& 0 & 1 & 0  \end{array} } \right\rangle 
- \left|\small{\begin{array}{cccccc} 0 & 0 & 0 & 0 & 0 & \cdot \\ \cdot & 1 & 1 & 0 & 1 & 1 \end{array}  } \right\rangle 
 - \left|\small{\begin{array}{cccccc} 1 & 0 & 0 & 0 & 1 & \cdot \\ \cdot & 0 & 1 & 0 & 1 & 0  \end{array} } \right\rangle
- \left|\small{\begin{array}{cccccc} 1 & 0 & 0 & 0 & 0 & \cdot \\ \cdot & 0 & 1 & 0 & 1 & 1 \end{array} }\right\rangle
\Bigg) s^- \ldots \Bigg] \\
& \quad - Q - (\rotatebox[origin=c]{-45}{$\leftrightarrow$}) + (Q, \rotatebox[origin=c]{-45}{$\leftrightarrow$}),
\end{aligned}
\end{equation}
with $Q$ the particle-hole conjugation operator and (\rotatebox[origin=c]{-45}{$\leftrightarrow$}) denoting interchanging the diagonals. Acting on the outer indices and projecting back to the $E=0$ space adds a new irreducible combination, named $\left| w_{ttt}\right\rangle$,
\begin{equation}
H'\left| w_{(1)}\right\rangle- \left|w_{(0)}\right\rangle= \sqrt{3 \over 2} \left| w_{(0)}'\right\rangle - {1 \over \sqrt{2}} \left| w_{ss}\right\rangle + \sqrt{2} \, \left| w_{ttt}\right\rangle
\end{equation}
with 
\begin{equation}
\begin{aligned}
\left| w_{ttt}\right\rangle = & {1 \over 2}
\left[ \ldots s^- \left(    \left|\small{\begin{array}{cccccc} 1 & 1 & 1 & 0 & 0 & \cdot \\ \cdot & 0 & 1& 1 & 0 & 1  \end{array} }\right\rangle\right. \right. 
- \left|\small{\begin{array}{cccccc} 0 & 0 & 0 & 1 & 1 & \cdot \\ \cdot & 1 & 0 & 0 & 1 & 0 \end{array} }\right\rangle 
+ \left|\small{\begin{array}{cccccc} 0 & 1 & 0 & 0 & 1 & \cdot \\ \cdot & 1 & 1 & 0 & 0 & 0  \end{array}}\right\rangle 
 - \left. \left.  \left|\small{\begin{array}{cccccc} 1 & 0 & 1 & 1 & 0 & \cdot \\ \cdot & 0 & 0 & 1 & 1 & 1 \end{array} }\right\rangle
\right) s^- \ldots \right].
\end{aligned}
\end{equation}

So we obtain 
\begin{equation}
\left| w_{(2)}\right\rangle= {1 \over 2} \left[\sqrt{3 \over 2} \left| w_{(0)}'\right\rangle - {1 \over \sqrt{2}} \left| w_{ss}\right\rangle + \sqrt{2} \, \left| w_{ttt}\right\rangle\right]
\end{equation}
and 
\begin{equation}
H' \left|w_{(1)}\right\rangle- \left|w_{(0)}\right\rangle = 2 \left| w_{(2)}\right\rangle .
\end{equation}

Some further algebra shows that
\begin{equation}\begin{aligned}
\left\langle w_{(2)}|H'|w_{(2)}\right\rangle &= -\frac{\sqrt{3}}{4} \, \left\langle w_{(0)}' | H' |w_{ss} \right\rangle \\
&= -\frac{\sqrt{6}}{2} \, \left\langle w_{(0)}' |H' |w_{ts+st}\right\rangle\\
&=-{1 \over \sqrt{2}},
\end{aligned}
\end{equation}

where we used that $\left\langle w_{(0)}'| w_{st+ts}\right\rangle ={1 / \sqrt{3}}$.  To this order, the FSA matrix becomes (assuming $L~>~4$)
\begin{equation}
M_{\rm FSA}^{(2)}= \left( \begin{array}{ccc} 2\sqrt{2} & 1 & 0 \\ 1  & 0 & 2 \\ 0 & 2 & {-1 \over \sqrt{2}} \end{array} \right)
\end{equation}
with largest eigenvalue $E_{(2)}=3.270...$.

\subsection{Finding $\left| w_{(3)}\right\rangle$}

Again, we act with $H'$ on $\left| w_{(2)}\right\rangle$ and subtract the previous vectors to determine
\begin{equation}
\left| u\right\rangle= H' \left| w_{(2)}\right\rangle- 2 \left| w_{(1)}\right\rangle + {1 \over \sqrt{2}} \left| w_{(2)}\right\rangle.
\end{equation}
We collect the following terms in $\left| u\right\rangle$:
\begin{itemize}
	\item
	terms with zero $t^\pm$ combine into ${3 \over 4}\left| w_{ss}\right\rangle$
	\item
	terms with a single $t^\pm$ give $\sqrt{3 \over 16}\left| w_{(0)}'\right\rangle -\left| w_{st+ts}\right\rangle = \sqrt{11 \over 16} \left| w_t'\right\rangle$, with
	\begin{equation}
	\begin{aligned}
	\left| w_t'\right\rangle={1 \over 4 \sqrt{11}} &
	\left[ \ldots s^- \left(  -  \left|\small{ \begin{array}{cccccc} 0 & 0 & 0 & 0 & 1 & \cdot \\ \cdot & 1 & 1& 0 & 1 & 0  \end{array} }\right\rangle \right. \right. 
+3 \left|\small{\begin{array}{cccccc} 0 & 0 & 0 & 0 & 0 & \cdot \\ \cdot& 1 & 1 & 0 & 1 & 1 \end{array}}\right\rangle 
	+3 \left|\small{\begin{array}{cccccc} 1 & 0 & 0 & 0 & 1 & \cdot \\\cdot & 0 & 1 & 0 & 1 & 0  \end{array}}\right\rangle 
	\left. \left. -5 \left|\small{\begin{array}{cccccc} 1 & 0 & 0 & 0 & 0 & \cdot \\ \cdot & 0 & 1 & 0 & 1 & 1 \end{array} }\right\rangle
	\right) s^- \ldots \right] \\
	& - (Q) - (\rotatebox[origin=c]{-45}{$\leftrightarrow$}) + (Q, \rotatebox[origin=c]{-45}{$\leftrightarrow$}).
	\end{aligned}
	\end{equation}
	\item
	terms with two adjacent $t^\pm$ give $\sqrt{9 \over 8} \left| w_{tt}'\right\rangle$, with
	\begin{equation}
	\begin{aligned}
	\left| w_{tt}'\right\rangle=& {1 \over \sqrt{8}}
	\left[ \ldots s^- s^+ \left(   \left|\small{\begin{array}{ccccc} 0 & 1 & 0 & 0 & \cdot \\ \cdot & 1 & 1& 0 & 1 \end{array} }\right\rangle \right. \right. 
	 -  \left|\small{\begin{array}{ccccc} 1 & 0 & 1 & 1 & \cdot \\ \cdot & 0 & 0 & 1 & 0  \end{array} }\right\rangle -  \left|\small{\begin{array}{ccccc} 1 & 1 & 1 & 0 & \cdot\\ \cdot & 0 & 1 & 1 & 1  \end{array}}\right\rangle	
	 \left. \left. +  \left|\small{\begin{array}{ccccc} 0 & 0 & 0 & 1 &  \cdot \\ \cdot & 1 & 0 & 0 & 0  \end{array} }\right\rangle
	\right) s^- \ldots \right]\\
	& \hspace{-3pt}+ {1 \over \sqrt{8}} \left[ \ldots s^- \left(   \left|\small{\begin{array}{ccccc} 0 & 1 & 0 & 0 & \cdot \\ \cdot & 1 & 1& 0 & 1 \end{array} }\right\rangle\right. \right.
	 -  \left|\small{\begin{array}{ccccc} 1 & 0 & 1 & 1 & \cdot \\ \cdot & 0 & 0 & 1 & 0  \end{array} }\right\rangle   +  \left|\small{\begin{array}{ccccc} 1 & 1 & 1 & 0 & \cdot \\ \cdot & 0 & 1 & 1 & 1  \end{array}}\right\rangle 
	 \left. \left.-  \left|\small{\begin{array}{ccccc} 0 & 0 & 0 & 1 &  \cdot \\ \cdot & 1 & 0 & 0 & 0  \end{array} }\right\rangle
	\right) s^+ s^- \ldots \right].
	\end{aligned}
	\end{equation}
	\item
	terms with two $t^\pm$ separated by an $s^\pm$ give ${1 \over \sqrt{2}} \left| w_{tst}\right\rangle$, with
	\begin{equation}
	\begin{aligned}
	\left| w_{tst}\right\rangle=& {1 \over 2}
	\left[ \ldots s^- \left(   \left|\small{\begin{array}{cccccc} 0 & 0 & 0 & 1 & 0 & \cdot \\ \cdot & 1 & 0 & 1 & 1 & 1 \end{array} }\right\rangle  \right. \right. 
-  \left|\small{\begin{array}{cccccc} 0 & 1 & 0 & 0 & 0 & \cdot \\ \cdot & 1& 1 & 1 & 0 & 1  \end{array}}\right\rangle 
-  \left|\small{\begin{array}{cccccc} 1 & 1 & 1 & 0 & 1 & \cdot \\ \cdot & 0 & 1 & 0 & 0 & 0  \end{array}}\right\rangle 
	\left.\left.+  \left|\small{\begin{array}{cccccc} 1 & 0 & 1 & 1 & 1 &  \cdot \\ \cdot & 0 & 0 & 0 & 1 & 0  \end{array} }\right\rangle
	\right) s^- \ldots \right].
	\end{aligned}
	\end{equation}
	\item
	terms with 3 adjacent $t^\pm$ give ${1 \over 2} \left| w_{ttt}\right\rangle$.
	\item
	terms with 4 adjacent $t^\pm$ give $\left| w_{tttt}\right\rangle$, with
	
	\begin{equation}
	\begin{aligned}
	\left| w_{tttt}\right\rangle=& {1 \over 2}
	\left[ \ldots s^- \left(   \left|\small{\begin{array}{ccccccc} 1 & 0 & 1 & 1 & 0 & 0 & \cdot \\ \cdot & 0 & 0 & 1 & 1 & 0 & 1 \end{array} }\right\rangle \right. \right.
	+  \left|\small{\begin{array}{ccccccc} 0 & 1 & 0 & 0 & 1 & 1 & \cdot\\ \cdot & 1 & 1 & 0 & 0 & 1 & 0  \end{array}}\right\rangle \\
	&\qquad + \left|\small{\begin{array}{ccccccc} 0 & 0 & 0 & 1 & 1 & 0 & \cdot \\ \cdot & 1 & 0 & 0 & 1 & 1 & 1  \end{array}}\right\rangle
	\left. \left. +  \left|\small{\begin{array}{ccccccc} 1 & 1 & 1 & 0 & 0 & 1 &  \cdot \\ \cdot & 0 & 1 & 1 & 0 & 0 & 0  \end{array} }\right\rangle
	\right) s^- \ldots \right].
	\end{aligned}
	\end{equation}
	
	The total result is
	\begin{equation}
	\begin{aligned}
	\left| u\right\rangle = &{3 \over 4}\left|w_{ss} \right\rangle + \sqrt{11 \over 16} \left|w_t' \right\rangle + \sqrt{9 \over 8} \left| w_{tt}'\right\rangle +
	 {1 \over \sqrt{2}} \left|w_{tst} \right\rangle + {1 \over 2} \left|w_{ttt} \right\rangle + \left|w_{tttt} \right\rangle
	\end{aligned}
	\end{equation}
	so that finally
	\begin{equation}
	\begin{aligned}
	H'\left| w_{(2)}\right\rangle = 2 \left| w_{(1)}\right\rangle- {1 \over \sqrt{2}} \left| w_{(2)}\right\rangle + \sqrt{33 \over 8} \left| w_{(3)}\right\rangle
	\end{aligned}
	\end{equation}
	with
	\begin{equation}
	\begin{aligned}
	\left| w_{(3)}\right\rangle = &\sqrt{8 \over 33} \left[ {3 \over 4}\left|w_{ss} \right\rangle + \sqrt{11 \over 16} \left| w_t'\right\rangle + \sqrt{9 \over 8} \left| w_{tt}'\right\rangle + {1 \over \sqrt{2}} \left|w_{tst} \right\rangle + {1 \over 2} \left| w_{ttt}\right\rangle + \left|w_{tttt} \right\rangle  \right].
	\end{aligned}
	\end{equation}
	
	It is useful to note that
	\begin{equation}
	\begin{aligned}
	& \left\langle w_{st+ts}|w_{(0)}' \right\rangle  ={1 \over \sqrt{3}}, &&\quad \left\langle w_{st+ts} | w_t' \right\rangle ={-3 \over \sqrt{11}},\\
	& \left\langle w_{(0)}'|w_t' \right\rangle ={-1 \over \sqrt{33}}, &&\quad\left\langle w_{tt}|w_{tt}' \right\rangle =0 .
	\end{aligned}
	\end{equation}
	With this one easily checks that all vectors $\{ \left|w_{(0)} \right\rangle, \left|w_{(1)} \right\rangle, \left|w_{(2)} \right\rangle, \left| w_{(3)}\right\rangle\}$ are orthonormal, as they should be. In order to calculate $\left\langle w_{(3)} |H'| w_{(3)} \right\rangle$, we make use of
	\begin{equation}
	\begin{aligned}
	& \left\langle w_{ss}| H' |w'_t \right\rangle = -{6 \sqrt{2} \over \sqrt{11}},
	&& \left\langle w_t'| H'| w'_{tt} \right\rangle={1 \over \sqrt{11}}, && \left\langle w_{tttt} |H'| w_{ttt} \right\rangle =\sqrt{2},\\
	& \left\langle w_t'| H'| w_{tst} \right\rangle ={2 \over \sqrt{11}},
	&& \left\langle w_{tt}'| H'| w_{ttt} \right\rangle =1.&&
	\end{aligned}
	\end{equation}
	This leads to
	\begin{equation}
	\left\langle w_{(3)} |H'| w_{(3)}  \right\rangle = {\sqrt{2} \over 11}\, .
	\end{equation}	
	To this order, the FSA matrix becomes (assuming $L>4$)
	\begin{equation}
	M_{\rm FSA}^{(3)}= \left( \begin{array}{cccc} 2\sqrt{2} & 1 & 0 & 0 \\ 1  & 0 & 2 & 0 \\ 0 & 2 & {-1 \over \sqrt{2}} & \sqrt{33 \over 8}
	\\ 0 & 0 & \sqrt{33 \over 8} & {\sqrt{2} \over 11}  \end{array} \right)
	\end{equation}
	with largest eigenvalue $E_{(3)}$ such that we got for the successive approximations $(E_{(1)},E_{(2)},E_{(3)})=(3.146\ldots,3.270\ldots,3.351\ldots)$. As can be seen the energies of the excited states keep increasing with higher orders of the FSA and are crucially not dependent on $L$, except that the FSA stops after a finite number of steps for every $L$. Continuing the FSA without cut off will lead to the energy of the excited state in the limit $L\rightarrow\infty$, that was also approximated in the main text by a fit to the energies of $L=4,6,8$ to be $E\approx3.49h_x$.
	
\end{itemize}

	\section{Entanglement Entropy of the $\ket{\psi_{E=0}}$ state}
	
	In this section we evaluate the entanglement entropy of the horizontal half-ladder by explicitly finding the reduced density matrix. In analogy to the left leaning singlets $s^-$, we define the right leaning singlets as 
	\beq
	r^- = \small{\left| \begin{array}{cc} \cdot & 0 \\ 1 & \cdot \end{array} \right\rangle- \left| \begin{array}{cc} \cdot & 1 \\ 0 & \cdot \end{array} \right\rangle}
	= \left|
	\begin{tikzpicture}[scale=\scale,baseline=\bs]
	\node at (0,1) {$\cdot$};
	\node at (1,0) {$\cdot$};
	\draw[black, thin] (0,0) -- (1,1);
	\end{tikzpicture}
	\right\rangle,
	\eeq
	where in the last equality we have introduced a graphical notation for the singlets. In this way we can write the density matrix of the symmetrized zero energy state as
	\beq
	\rho = \ket{\mathcal R}\bra{\mathcal R} + \ket{\mathcal R}\bra{\mathcal S} + \ket{\mathcal S}\bra{\mathcal R} + \ket{\mathcal S}\bra{\mathcal S}.
	\label{eq:rho}
	\eeq
	Here
	\begin{subequations}
		\label{eq:R_n_S}
		\begin{align}
		\ket{\mathcal R} &= \Ket{\rb \ldots \rb}, \\	
		\ket{\mathcal{S}} &= \Ket{\lb \ldots \lb}, 
		\end{align}
	\end{subequations}
	where the periodic boundary conditions are assumed on the last singlet, which connects sites with $x$-coordinate $L$ and 1. In the following we refer to sites simply by their $x$-coordinate, unless stated otherwise. Since we consider only even lengths of the ladder, we parametrize it as $L=2l$. We divide the ladder into two subsystems by breaking the singlets between sites $l,l+1$ and $L,1$, such that the states (\ref{eq:R_n_S}) can be written as
	\begin{subequations}
		\begin{align}
		\ket{\mathcal R} &= -\ket{{}^{00}R}\ket{{}^{11}R'}+\ket{{}^{01}R}\ket{{}^{01}R'}+\ket{{}^{10}R}\ket{{}^{10}R'}-\ket{{}^{11}R}\ket{{}^{00}R'}, \\
		\ket{\mathcal S} &= -\ket{{}^{00}S}\ket{{}^{11}S'}+\ket{{}^{01}S}\ket{{}^{01}S'}+\ket{{}^{10}S}\ket{{}^{10}S'}-\ket{{}^{11}S}\ket{{}^{00}S'},
		\end{align}
	\end{subequations}
	where we have labeled the right subsystem with a prime and introduced a notation
	\begin{subequations}
		\begin{align}
		\ket{{}^{xy}R} &= \Ket{\rstatevar{x}{y}}, \\
		\ket{{}^{xy}S} &= \Ket{\lstatevar{x}{y}}.
		\end{align}
	\end{subequations}
	In order to perform the trace over the primed subsystem of (\ref{eq:rho}), we first need to find a suitable basis. To this end we note, that while $\ket{{}^{xy}R'}$ and $\ket{{}^{x'y'}R'}$ are orthogonal for all $x \neq x'$, $y \neq y'$, $\ket{{}^{xy}S'}$ and $\ket{{}^{x'y'}R'}$ are not. In other words, we are looking for a suitable decomposition of $\ket{{}^{xy}S'}$ on $\ket{{}^{x'y'}R'}$. To proceed, we will distinguish two cases, $l$ even and $l$ odd.

	\subsection{$l$ odd}
	
	To find the decomposition of  $\ket{{}^{xy}S'}$ on $\ket{{}^{x'y'}R'}$, we note that two sites belonging to a singlet can feature occupation numbers (0,1) or (1,0), but never (0,0) or (1,1). Writing a specific example of
	\begin{subequations}
		\begin{align}
		\ket{{}^{00}S'} &= \Ket{\lstatevar{0}{0}}, \\
		\ket{{}^{xy}R'} &= \Ket{\rstatevar{x}{y}},
		\end{align}
	\end{subequations}
	we see that the state components of $\ket{{}^{xy}R'}$ with non-zero overlap with $\ket{{}^{00}S'}$ have to contain (0,1) on the first right leaning singlet
	\[
	\Ket{
		\begin{tikzpicture}[scale=0.5,baseline=0.15cm]
		\node at (1,0.2) {0};
		\node at (1.8,0.8) {1};
		\node at (3.5,0.5) {\ldots};
		\draw[black, thin] (2,0) -- (3,1);
		\draw[black, thin] (4,0) -- (5,1);
		\end{tikzpicture}
	}.
	\]
	The presence of the occupation ``1'' then forces the configuration (1,0) on the first left-leaning singlet and so on, until we reach the right end which fixes $y=0$. Performing the same procedure starting from the right, which fixes $x=0$, we find that the only component of $\ket{{}^{xy}R'}$ with non-vanishing overlap with $\ket{{}^{00}S'}$ is the $\left| \mathbb{Z}_2\right\rangle$ state, so that we can write
	\beq
	\ket{{}^{00}S'} = \frac{1}{\sqrt{w}} \left(\ket{{}^{00}R'} + \sqrt{w-1} \ket{{}^{00}R'_\perp} \right),
	\label{eq:S00}
	\eeq
	where $\ket{{}^{00}R'_\perp}$ is the orthogonal complement of $\ket{{}^{00}R'}$, $\Braket{{}^{00}R'|{}^{00}R'_\perp}=0$. Importantly, while so far we have considered unnormalized states, it is now necessary to include the proper normalization in order to get the correct structure for the reduced density matrix. Expanding $\ket{{}^{00}S'}$ in the basis states yields the coefficient of each basis state of magnitude $1/2^{(l-1)/2}$ and similarly for $\ket{{}^{00}R'}$, so that $\Braket{{}^{00}R'|{}^{00}S'}=1/2^{l-1}$. At the same time, from (\ref{eq:S00}) we have $\Braket{{}^{00}R'|{}^{00}S'}=1/\sqrt{w}$, which implies $w=2^{2(l-1)}$.
	
	Continuing the same procedure for the remaining $S'$ states yields
	\begin{subequations}
		\label{eq:S}
		\begin{align}
		\ket{{}^{01}S'} &= \frac{1}{\sqrt{w}} \left(\ket{{}^{10}R'} + \sqrt{w-1} \ket{{}^{10}R'_\perp} \right), \\
		\ket{{}^{10}S'} &= \frac{1}{\sqrt{w}} \left(\ket{{}^{01}R'} + \sqrt{w-1} \ket{{}^{01}R'_\perp} \right), \\
		\ket{{}^{11}S'} &= \frac{1}{\sqrt{w}} \left(\ket{{}^{11}R'} + \sqrt{w-1} \ket{{}^{11}R'_\perp} \right).
		\end{align}
	\end{subequations}
	Since $\Braket{{}^{xy}S'|{}^{x'y'}S'}=\delta_{xx'}\delta_{yy'}$, we conclude that the set $\{\ket{{}^{00}R'},\ket{{}^{01}R'},\ket{{}^{10}R'},\ket{{}^{11}R'},\ket{{}^{00}R'_\perp},\ket{{}^{01}R'_\perp},\ket{{}^{10}R'_\perp},\ket{{}^{11}R'_\perp}\}$ constitutes the necessary orthonormal basis (together with its non-primed counterpart) for the decomposition of $\rho$. With the relations (\ref{eq:S00}),(\ref{eq:S}) we can now perform the partial trace of (\ref{eq:rho}) and after some algebra we find an expression for the reduced density matrix of the half-ladder in the non-primed basis
	\beq
	\rho_{\rm red}=
	\left(
	\begin{array}{cccccccc}
		1+\frac{3}{w} & 0 & 0 & 0 & \frac{2 \sqrt{w-1}}{w} & 0 & 0 & 0 \\
		0 & 1+\frac{3}{w} & 0 & 0 & 0 & \frac{2 \sqrt{w-1}}{w} & 0 & 0 \\
		0 & 0 & 1+\frac{3}{w} & 0 & 0 & 0 & \frac{2 \sqrt{w-1}}{w} & 0 \\
		0 & 0 & 0 & 1+\frac{3}{w} & 0 & 0 & 0 & \frac{2 \sqrt{w-1}}{w} \\
		\frac{2 \sqrt{w-1}}{w} & 0 & 0 & 0 & \frac{w-1}{w} & 0 & 0 & 0 \\
		0 & \frac{2 \sqrt{w-1}}{w} & 0 & 0 & 0 & \frac{w-1}{w} & 0 & 0 \\
		0 & 0 & \frac{2 \sqrt{w-1}}{w} & 0 & 0 & 0 & \frac{w-1}{w} & 0 \\
		0 & 0 & 0 & \frac{2 \sqrt{w-1}}{w} & 0 & 0 & 0 & \frac{w-1}{w}
	\end{array}
	\right).	
	\eeq
	As we are interested in the evaluation of the second R\'{e}nyi entanglement entropy, we find (with the overall normalization)
	\beq
	\frac{{\rm Tr}\left( \rho_{\rm red}^2\right)}{{\rm Tr}\left(\rho_{\rm red} \right)^2} = \frac{1+w(6+w)}{8(1+w)^2}.
	\eeq

	\subsection{$l$ even}
	
	Proceeding along similar lines as in the $l$-odd case, we first find that $\Braket{{}^{00}S'|{}^{xy}R'}=\Braket{{}^{11}S'|{}^{xy}R'}=0$ so that $\ket{{}^{00}S'},\ket{{}^{11}S'}$ are part of the basis on which we seek to decompose the $\ket{\psi_{E=0}}$ state. Next, we find 
	\begin{subequations}
		\label{eq:Seven}
		\begin{align}
		\ket{{}^{01}S'} &= \frac{1}{\sqrt{w}} \left(-\ket{{}^{01}R'} + \ket{{}^{01}R'} + \sqrt{w-2} \ket{{}^{01} R'_\perp} \right), \\
		\ket{{}^{10}S'} &= \frac{1}{\sqrt{w}} \left(\ket{{}^{01}R'} - \ket{{}^{10}R'} + \sqrt{w-2} \ket{{}^{10}R'_\perp} \right),
		\end{align}
	\end{subequations}
	from where it follows that
	\beq
	\Braket{{}^{01}R'_\perp|{}^{10}R'_\perp} = \frac{2}{w},
	\eeq
	which forces us to further decompose $\ket{{}^{01}R'_\perp}, \ket{{}^{10}R'_\perp}$ as a sum of a state we denote as $\ket{\delta R'}$, which is common to both, and the remainders $\ket{{}^{01}\Delta R'}, \ket{{}^{10} \Delta R'}$. The Eqs. (\ref{eq:Seven}) become
	\begin{subequations}
		\begin{align}
		\ket{{}^{01}S'} &= \frac{1}{\sqrt{w}} \left(-\ket{{}^{01}R'} + \ket{{}^{01}R'} + \sqrt{2}\ket{\delta R'}+\sqrt{w-4} \ket{{}^{01} \Delta R'} \right), \\
		\ket{{}^{10}S'} &= \frac{1}{\sqrt{w}} \left(\ket{{}^{01}R'} - \ket{{}^{10}R'} + \sqrt{2}\ket{\delta R'}+ \sqrt{w-4} \ket{{}^{10} \Delta R'} \right).
		\end{align}
	\end{subequations}
	We thus have the orthonormal basis $\{\ket{{}^{00}R},\ket{{}^{01}R},\ket{{}^{10}R},\ket{{}^{11}R},\ket{\delta R},\ket{{}^{01}\Delta R'},\ket{{}^{10}\Delta R'},\ket{{}^{00}S},\ket{{}^{11}S}\}$ in which the reduced density matrix is evaluated to
	\beq
	\rho_{\rm red}=
	\left(
	\begin{array}{ccccccccc}
		1 & 0 & 0 & 0 & 0 & 0 & 0 & 0 & 0 \\
		0 & 1+\frac{6}{w} & -\frac{6}{w} & 0 & 0 & -\frac{2 \sqrt{w-4}}{w} & \frac{2 \sqrt{w-4}}{w} & 0 & 0 \\
		0 & -\frac{6}{w} & 1+\frac{6}{w} & 0 & 0 & \frac{2 \sqrt{w-4}}{w} & -\frac{2 \sqrt{w-4}}{w} & 0 & 0 \\
		0 & 0 & 0 & 1 & 0 & 0 & 0 & 0 & 0 \\
		0 & 0 & 0 & 0 & \frac{4}{w} & \frac{\sqrt{2} \sqrt{w-4}}{w} & \frac{\sqrt{2} \sqrt{w-4}}{w} & 0 & 0 \\
		0 & -\frac{2 \sqrt{w-4}}{w} & \frac{2 \sqrt{w-4}}{w} & 0 & \frac{\sqrt{2} \sqrt{w-4}}{w} & \frac{w-4}{w} & 0 & 0 & 0 \\
		0 & \frac{2 \sqrt{w-4}}{w} & -\frac{2 \sqrt{w-4}}{w} & 0 & \frac{\sqrt{2} \sqrt{w-4}}{w} & 0 & \frac{w-4}{w} & 0 & 0 \\
		0 & 0 & 0 & 0 & 0 & 0 & 0 & 1 & 0 \\
		0 & 0 & 0 & 0 & 0 & 0 & 0 & 0 & 1
	\end{array}
	\right)
	\eeq
	and
	\beq
	\frac{{\rm Tr}\left( \rho_{\rm red}^2\right)}{{\rm Tr}\left(\rho_{\rm red} \right)^2} = \frac{4+w(6+w)}{8(1+w)^2}.
	\eeq
	Finally, we remark that in the thermodynamic limit $l \rightarrow \infty$, the second R\'{e}nyi entanglement entropy evaluates to $3 \ln(2)$ for both $l$ even and odd.

\section{Counting of peak states}
In this section we show that the number of peak states is of the order $\phi^{2L}$ where $\phi=(1+\sqrt{5})/2\approx 1.618$ is the golden ratio.
A peak state on the Ising ladder is a basis state that is only connected by $H_x$ with basis states that have a different eigenvalue of $H_z$, i.e. the potential $V$. Since $H_x$ only changes the value of one occupation number at a time, this means that flipping any spin in the basis state changes the potential. The only local configuration that conserves the potential is when the number of equal and unequal neighbours is the same and is given in Eq.~(\ref{zeroflip}), which we here repeat for reader's convenience
\begin{equation}
\label{zeroflip_app}
\small{ \left| \begin{array}{ccccc} \cdots & n & \underline{n_a} & n & \cdots\\  \cdots & \cdot & 1-n & \cdot & \cdots \end{array} \right\rangle} \nonumber.
\end{equation}
We can then raise the question: given a ladder of length $L$, how many basis states exist that do not have the configuration Eq.~(\ref{zeroflip}) anywhere? The important quantity to look at are the sites that are diagonally placed from each other, i.e. the next-nearest neighbour sites. For an easier analysis we denote this chain of occupation numbers as one string as follows

\begin{equation}\label{stringmap}
\small{ \left| \begin{array}{ccccc} \cdots & n_{i,1} & \cdot & n_{i+2,1} & \cdots\\  \cdots & \cdot & n_{i+1,0} & \cdot & \cdots \end{array} \right\rangle}\; \rightarrow \;\cdots n_{i,1}^{\phantom{\dagger}}\ n_{i+1,0}^{\phantom{\dagger}}\ n_{i+2,1}^{\phantom{\dagger}}\cdots.
\end{equation}

Each basis state consists of two non-overlapping strings, one that starts at $n_{0,0}$ and one starting at $n_{0,1}$. For a basis state to lead to a peak state, both strings should not contain the substrings $101$ and $010$, because we want to exclude the configurations (\ref{zeroflip}). This leads to the constraint that in the string after a $10$, the next site is not allowed to be a $1$ and similarly after a $01$ the next one is not allowed to be a $0$. We can use these constraints to calculate the total number of allowed strings by making use of a transfer matrix. This matrix shows which configuration on sites $i+1$ and $i+2$ are allowed, based on the occupation of sites $i$ and $i+1$. Taking as basis states ($00, 01, 10, 11$), the transfer matrix is
\begin{equation}\label{transfer_mat}
T = \left( \begin{matrix}
1 & 1 & 0 & 0 \\ 
0 & 0 & 0 & 1 \\ 
1 & 0 & 0 & 0 \\ 
0 & 0 & 1 & 1
\end{matrix} \right).
\end{equation}
Because periodic boundary conditions are assumed, applying the transfer matrix $L$ times should return to the initial values of site 0 and 1. Therefore, the total number of allowed configurations of a string is $\text{Tr}(T^L)$, because the diagonal entries are the number of possibilities that a valid string starts and ends with the same contribution. The eigenvalues of (\ref{transfer_mat}) are $(1\pm\sqrt{5})/2,{\rm e}^{\pm i \pi/3}$ so that for large $L$ $\text{Tr}(T^L)\approx \phi^L$ will be dominated by the maximum magnitude eigenvalue, which is $\phi=(1 + \sqrt{5})/2$. Because every basis state consists of two of those strings, the number of peak states scales as $\phi^{2L}$.\\

\end{document}